\documentclass[12pt]{iopart}

\usepackage{graphicx}

\begin{document}

\title[Synge's dynamic problem for two isolated point charges]{Synge's dynamic problem for two isolated point charges. A new method to find global solutions for Functional Differential Equations System}

\author{Rodrigo R. Silva \& Annibal Figueiredo}
\address{Institute of Physics, University of Brasília, Brasília-DF, Brazil, 70919-970}
\ead{rodrigotdk@gmail.com.br \&  annibalfig@unb.br}
\vspace{10pt}

\begin{abstract}
Synge's problem consists in to determine the dynamics of two point electrical charges interacting through their electromagnetic fields, without to take into account the radiation terms due to the self-forces in each charge. 
We discuss how this problem is related to the question on to establish initial conditions for the electromagnetic fields that are compatible with the two point charges system isolation, that is, the charges are free from the action of external forces. This problem stems from the existence of  inter-temporal constraints for the charges trajectories, which implies that the relativistic Newton equations for the charges is not a system of ODEs, but rather a system of Functional Differential Equations (FDEs). 
We developed a new method to obtain global solutions that satisfies this system of FDEs and a given initial condition for the charges positions and velocities. This method allows the construction of  a recursive numerical algorithm that only use integration methods for ODEs systems.  
Finally, we apply this algorithm to obtain numerical approximations for the quasi-circular solutions that are predicted in  Synge's problem.
\end{abstract}


%
%
%
%
%

\section{Introduction}

The problem of two isolated point charges consists in to determine their trajectories from the forces generated only by their respective electromagnetic fields. The first rigorous formulation of this problem was made by Synge \cite{synge1940}, where he does not consider the self-force in each point charge \cite{lorentz1904,abraham1905,dirac1938,infeld1940}. Synge's mathematical problem consists in to solve Maxwell's equations  coupled with the  relativistic equations for the Lorentz force acting on each charge.
This is a very complex mathematical problem that,
in addition to the technical difficulties due to charge density singularities,  involves serious questions to establish an initial condition \cite{deckert2016}.
The origin of this difficulty relates to the existence of an inter-temporal constraint:
the Lienard-Wiechert potentials that allows to obtain the electromagnetic field due to a point charge with known trajectory \cite{smith2014,rohrlich2007}.

The problem with a given initial condition consists in to know if the initial state for the particles and  electromagnetic fields are compatible with the two-bodies system isolation, that is, if the charges are not under the influence of  other sources of electromagnetic fields.
Although this kind  of initial value problem can be formulated, it implies extremely complicated mathematical conditions for the electromagnetic fields at the initial time, without referring to the previous dynamics of the charges. This is actually a consequence of Lienard-Wiechert fields implying an inter-temporal constraint on charge trajectories. 
Another issue is the physical possibility of carrying out previous trajectories compatible with viable external forces that no longer influence the movement of charges after the initial instant, especially if these forces are related to other electromagnetic field sources.  For a good discussion on these problems, we recommend the reference \cite{deckert2016}.

Synge consider the explicit expressions for the retarded electromagnetic fields derived from the Lienard-Wiechert potentials and substitute them in the Lorentz force that appear in Newton's relativistic equations. 
These equations do not form an Ordinary Differential Equations (ODEs) system, but rather a Functional Differential Equations (FDEs) system, since they relate the acceleration of the charges at the present time with their positions, velocities and accelerations at  some past times \cite{myskis1951,driverbook1977,diekmann1995}. This type of FDE system is called neutral, because the differential equations relate the higher order derivatives calculated at different instants of time.

In his seminal work \cite{synge1940}, Synge had the great merit of alerting to the mathematical consistency of the problem formulated in terms of a system of FDEs, showing that this mathematical consistency is not correlated with issues associated to the conservation of momentum and energy for the complete electromagnetic system made up of charges and fields. Indeed, these issues depend exclusively on the way as  electromagnetic momentum and energy are defined and, from a mathematical point of view, are irrelevant to the obtention of the corresponding FDEs system.
The history of mathematical progress in characterizing solutions to Synge's problem, which is closely linked to the development of existence (and often uniqueness) theorems in neutral FDEs systems, confirms his warning
\cite{driver1963,driver1967,driver1969,schild1963,schild1968,andersen1972,andersen1972a,zhdanov1976,travis1975,hsing1977,driver1979,driver1984,
angelov1990,angelov2000,deckert2013,deckert2015} .

Therefore, the Synge problem consists in to search for charges trajectories that satisfy the FDEs system together with some initial condition of Cauchy type. In fact, this problem concerns the existence of global solutions and in the literature it is known as ``Backward Problem''\cite{driver1969}. This problem has not yet been solved in general, but analytical or numerical solutions are known for certain particular one-dimensional problems \cite{driver1969,travis1975,zhdanov1976,hsing1977,driver1979,deckert2015,franklin2014}. 
An even more restricted problem is whether for a given initial condition there is a global unique solution \cite{driver1969,zhdanov1976,hsing1977}. An existence but not uniqueness theorem for one-dimensional systems is derived in \cite{travis1975,deckert2015}. 

Furthermore, the Synge problem formulation for retarded electromagnetic fields can be generalized for advanced electromagnetic fields or any linear combination of retarded and advanced fields.
Indeed, by a linear combination of retarded and advanced fields,
we understand an electromagnetic field ${\cal E}=a{\cal E}_r+b{\cal E}_a$, where ${\cal E}_r$ denotes the retarded electromagnetic field and ${\cal E}_a$ the advanced electromagnetic field, with $a+b=1$ and $a,b\geq 0$. For the particular case where $a=b=1/2$, the electromagnetic field is called symmetric.
In  regarding the existence of global solutions for symmetric fields, it was shown the existence of circular, quasi-circular and some other particular solutions \cite{schild1963,andersen1972,andersen1972a}. Furthermore, existence and uniqueness theorems of global solutions, associated with a certain class of initial conditions, were demonstrated for one-dimensional  systems  \cite{driver1979}.

Still in relation to the work of Synge \cite{synge1940}, it is worth remembering that he develops a self-consistent iterative method that would allow obtaining solutions for a given initial condition. Essentially, this method consists of considering a given trajectory of one charge and using its Lienard-Wiechert fields to solve the Newton equation associated with the other charge.
From this trajectory obtained for the other charge, we repeat the procedure and recalculate the trajectory for the first one. This iterative procedure must be repeated indefinitely until it converges to the solution of the problem.
The convergence of this method would require that the masses of the point charges are different. To fix the initial trajectory Synge considers the Keplerian problem (analyzed by Sommerfeld in \cite{sommerfeld1934}), where the mass of one of the charges is considered infinite and, therefore, it is possible to have a well-defined ODEs system. From the second step onwards, a finite value for ratio $\eta$ between the mass must be considered.

Unfortunately, Synge did not rigorously demonstrate the convergence of  his method, which was only suggested heuristically. It is interesting to note that the Synge method has only been used successfully for cases where $\eta=1$ \cite{andersen1972,travis1975,franklin2014}.
In the few numerical simulations using his method for $ \eta > 1 $ there is no convergence, even for opposite charges with small velocities  in quasi-circular orbits \cite{andersen1972a}.
This fact is quite frustrating because it is in the attractive case that Synge develops approximate analytical formulas (which depend on $\eta$ being very large) to obtain the description of quasi-circular singular trajectories, that is, spiral trajectories of two charges that after a  finite time collide. The interesting point here is that the collision of the two charges would not be a consequence of  self-forces  radiations, since these are absent in the formulation of Synge's problem.

The work developed in \cite{angelov2003} criticizes Synge's iterative method showing that it could not be implemented after the first step. This criticism seems to point more to the question of the initial trajectory, chosen from a Keplerian problem with $\eta=\infty$ and its subsequent passage to a system with finite $\eta$, than a general criticism of the method regardless the chosen initial trajectory for one of particles. This fact is well demonstrated for the one-dimensional cases where the Synge approximation method was successfully used.

Another important observation is the existence of few numerical simulations of Synge's problem: either through implementations of his algorithm \cite{andersen1972a,franklin2014} or through the implementation of another type of numerical integration algorithms for FDEs systems \cite{kasher1971, huschilt1973}. This fact contrasts sharply with the abundance of numerical methods developed to calculate the radiation field generated by the trajectory of a single point charge with a known trajectory. In fact, this problem is very important in analysing the radiation of a charge in a particle accelerator \cite{ryne2018,mayes2021}.

The great difficulty associated with FDEs system  is to build global solutions from local solutions, as is the case for ODEs system. Indeed, the very notion of a local solution does not exist due to the presence of functional equations relating the time dependent variables calculated at different instants. 
The main contribution of this work is to develop a method of constructing global solutions, starting from a given initial condition, which in a certain way circumvents the previously mentioned difficulty.
We developed a method to construct a sequence of ODEs systems, such that for a given initial condition, if the sequence of solutions of the respective ODE systems converges to a well-determined function, then this function will be a solution of the Synge's FDEs system.
Thus, these solutions can be seen as successive approximations of the global solution to the problem. 
The difference to Synge's original method is that the trajectories of the two charges are obtained simultaneously for each system of ODEs in the sequence. It is worth noting that the method is self-consistent, since an ODEs system is built from the solution of the previous system in the sequence. 
Moreover, we developed a numerical integration algorithm to obtain  the solution for each ODEs system in the sequence constructed by the method and, in this way, we show evidence of convergence for quasi-circular trajectories: spiral trajectories for systems with retarded fields and oscillating trajectories for systems with symmetric fields.

It is important to draw the atention to a weaker formulation of the Synge's problem, made in two seminal works by Driver \cite{driver1963,driver1967}, that is based on typical formulation for the question of solution existence and uniqueness for retarded systems:  the search not for global solutions for all time $t$, but just solutions for  $t>t_0$, where $t_0$ is a given initial instant.
In this weaker formulation, the proof of an existence and uniqueness theorem is associated with the establishment of some conditions
for the trajectories in a certain finite time interval before the initial time $t_0$. This condition replaces the typical Cauchy initial condition for ODEs problems, where the trajectory is known only at the initial time $t_0$.
Driver proves, with slightly restrictive hypotheses about the differentiability of the previous trajectories together with the hypothesis that the functional equations (for the retarded times) are satisfied at $t_0$, a very general theorem for the existence and uniqueness of solutions.
In another work \cite{driver1960}, much less known in the literature, Driver obtains the same theorem for the three-dimensional problem. The results of this work were republished later with simpler proofs and hypotheses \cite{driver1984}. Even so, several later works ignore this work and get Driver's results again under more restrictive conditions \cite{angelov1990,angelov2000,murdock1973,murdock1979}.

The theorem obtained by Driver solves, in a very general way, the problem of existence and uniqueness for the weaker formulation of Synge's problem. However, Driver himself, in another work \cite{driver1969}, draws attention to the fact that solving the weaker formulation does not solve the problem as originally formulated by Synge. Due to the existence and uniqueness theorem for the weaker formulation,  to solve Synge's original problem it is enough to obtain trajectories that satisfy the FDEs system for all time $t<t_0$. For this reason, the problem of finding global solutions became known as the ``Backward Problem''. 

We emphasize that the solutions obtained from this weaker formulation do not guarantee the isolation of the two charges for times prior to the time $t_0$, implying that they would be under the influence of external fields. Indeed, to  consider arbitrary trajectories for the charges prior the time $t_0$,  it would be necessary to suppose external fields that, togheter with the fields generated by the charges, would determine these trajectories.

The weaker version of Synge's three-dimensional problem for symmetric fields was demonstrated in references \cite{deckert2013} and \cite{deckert2015}, where the idea of knowing a previous trajectory is replaced by the knowledge of asymptotic trajectories in the intervals $(- \infty,- T)$ and $(T,\infty)$ for some $T>0$. In this case, an existence and uniqueness theorem guarantees the existence of a unique trajectory in the finite interval $[-T,T]$. From this result, the author demonstrates a theorem similar to the one obtained by Driver \cite{driver1984} for Synge's original problem.  
Furthermore,  they indicate that the proofs for the general case with arbitrary linear combinations between the retarded and advanced fields would be analogous. As far as we know, this seems to be the only existence and uniqueness theorem for the general case of linearly combined retarded and advanced electromagnetic fields. 

Another  important issue that must be pointed out is that the Synge problem for retarded fields considers two point charges without taking into account the reaction forces arising from the self-interaction of each charge on itself, called radiation forces and originally obtained by Lorentz and Abraham \cite{ lorentz1904, abraham1905}. In order to obtain this self-force they considered the forces between the parts of a small rigid sphere (uniformly charged), taking its volume to zero and keeping the total amount of charge. This procedure is  known as renormalization and was obtained in \cite{dirac1938,infeld1940} without using the idea of a spheric rigid charge.

The problem of two point charges with self-interaction can be formulated in analogous way as the generalized Synge problem, but taking into account the correction of  the Newton's equation, leading to the appearance of first-order derivatives for the charge accelerations. For retarded fields, the FDEs system obtained is a non-neutral third-order differential system and is currently called Dirac-Lorentz equations. The renormalization process for symmetric fields leads to self-forces at retarded times that cancel out the self-forces at advanced times, and no correction is necessary for the generalized Synge problem. These neutral second-order FDEs are called Wheeller-Feynman equations  \cite{feynman1945,feynman1949}. A good critical discussion on the Dirac-Lorentz versus Wheeler-Feyman equations can be seen in \cite{havas1948}.

We observe that to consider Dirac-Lorenz equations instead of  Synge equations, with the respective self-forces corrections for each charge, does not profoundly change the mathematical problem of finding global solution and the respective problem to assure the system isolation from external fields.
However, from a certain point of view, we can consider only the Synge's problem as a true pure electromagnetic problem. The self-force corrections for each charge in the Dirac-Lorentz equations can be derived by renormalizing rigid bodies occupying a certain volume, then the interaction between two points in this volume can be view as a Synge problem with the distance between the points fixed \cite{beil1975}, which certainly means to suppose an external non-electromagnetic field to hold this distance.

\section{Functional Differential Equations of the Synge Problem}

Synge's mathematical problem consists in to couple Maxwell's equations for two point charges with the relativistic mechanics equations using the Lorentz Force on each charge.
Newton's relativistic equations for two accelerated point charges are given by:
\begin{equation}
\frac{d{\mathbf{p}_{i}}}{dt} = \mathbf{F}_{ji},\;\; 
\mathbf{p_{i}} = \frac{m_{i}\mathbf{v}_{i}}{\sqrt{1-|\mathbf{v}_{i}|^2/c^2}},\; i=1,2,
\label{newton_tr_ta}
\end{equation}
where $\mathbf{p}_i$ is the relativistic moment of the particle $i$,
\begin{equation}
\label{forcadelorentz}
\mathbf{F}_{ji} =  q_{i}\,\mathbf{E}_{ji}+ q_{i}\,\mathbf{v}_{i} \times \mathbf{B}_{ji}\;\;(i\neq j)
\end{equation}
is the Lorenz force on the point charge $i$ due to the electromagnetic field of the point charge $j$, and ${\bf v}_i$ is the velocity of the particle $i$. We are considering the charge $q_i$ as a point particle with inertial mass $m_i$.

The electromagnetic fields  satisfy the Maxwell's equations for a distribution of two point charges, which corresponds to a system of  Partial Differential Equations (PDEs) given by:
\begin{equation}
\label{maxwell}
\begin{array}{lll}
\nabla\cdot {\bf B}\left({\bf r},t\right)=0, &&\displaystyle  \nabla\times{\bf E}\left({\bf r},t\right)+\frac{\partial {\bf B}\left({\bf r},t\right)}{\partial t}=0, \\ &\\
\displaystyle \nabla {\bf E}\left({\bf r},t\right)=\frac{\rho\left({\bf r},t\right)}{\epsilon_0}, &&\displaystyle  \nabla\times{\bf B}\left({\bf r},t\right)-\frac{1}{c^2}\frac{\partial {\bf E}\left({\bf r},t\right)}{\partial t}=\mu_0{\bf j}\left({\bf r},t\right), \\
\end{array}
\end{equation}
where ${\bf r}$ is a position vector in  the tridimensional space and $t$ an instant of time. The  Maxwell equations are written in the MKS system of units, with $\epsilon_0$, $\mu_0$ and $c$ being respectively  the electrical permittivity, magnetic susceptibility and speed of light in the vacuum. 
The charge densities $\rho({\bf r},t)$ and current densities ${\bf j}({\bf r},t)$ are defined as:
\begin{equation}
\label{densidades}
\rho({\bf r},t)=\sum_{i=1,2}q_i\delta({\bf r}-{\bf r}_i(t)),\;\;{\bf j}({\bf r},t)=\sum_{i=1,2}q_i{\bf v}_i(t)\delta({\bf r}-{\bf r}_i(t)),
\end{equation}
where $\delta$ is the three-dimensional $\delta$-Dirac  function and ${\bf r}_i$ is the position vector of the charge $q_i$. The electromagnetic fields  in (\ref{forcadelorentz}) should be calculated at the charge position ${\bf r}_i$ and instant $t$:
\begin{equation}
\label{campos}
{\bf E}_{ji}={\bf E}({\bf r}_i,t),\;\;{\bf B}_{ji}={\bf B}({\bf r}_i,t)
\end{equation} 

The  Maxwell's equations (\ref{maxwell}) has a well-determined solution, builted from the  Lienard-Wiechert potentials, for a given charge trajectory.
Indeed, the retarded or advanced Lienard-Wiechert fields are a time-varying electromagnetic fields satisfying the Maxwel equations for a point charge $q$ in an arbitrary motion described by its position vector ${\bf r}_q(t)$ as a function of time $t$.

The respective  Liénard-Wiechert retarded electromagnetic fields are given by \cite{smith2014}:
\begin{equation}
\begin{array}{l}
\displaystyle \mathbf{E}^-({\bf r},t) = \frac{q}{4 \pi \epsilon_{0}}\left[\frac{(1-\beta^{2}
)(\mathbf{n}-{\beta})}{{R}^{2}|1-\mathbf{n}\cdot{\beta}|^{3}} + \frac{\mathbf{n}\times[(\mathbf{n}-{\beta})\times\mathbf{a}]}{c^{2}R|1- \mathbf{n}\cdot{\beta}|^{3}}\right]_{t^r},\\ \\
\displaystyle \mathbf{B}^-({\bf r},t)= \frac{\mathbf{n}(t^r)}{c} \times \mathbf{E}^-({\bf r},t),\;\;\;t^r = t - \frac{R(t^r)}{c},
\label{E_Field_tr}
\end{array}
\end{equation}
where ${\bf R}={\bf r}-{\bf r}_q(t)$, $R=|{\bf R}|$, $\mathbf{n} = \mathbf{R}/R$, ${\beta} = \mathbf{v}/c$, ${\bf v}=d{\bf r}_q/dt$ and ${\bf a}=d{\bf v}_q/dt$ are all evaluated at
the retarded time $t^r$.
The derivation of Liérnard-Wiechert advanced fields is similar to the derivation of retarded fields, where we consider the electromagnetic fields evaluated at the advanced time $t^a$:
\begin{eqnarray}
\begin{array}{l}
\displaystyle \mathbf{E}^+ ({\bf r},t)= \frac{q}{4 \pi \epsilon_{0}}\left[\frac{(1-\beta^{2}
)(\mathbf{n}+{\beta})}{{R}^{2}|1+\mathbf{n}\cdot{\beta}|^{3}} + \frac{\mathbf{n}\times[(\mathbf{n}+{\beta})\times\mathbf{a}]}{c^{2}R|1+\mathbf{n}\cdot{\beta}|^{3}}\right]_{t^a},\\ \\
\displaystyle \mathbf{B}^+({\bf r},t)= \frac{\mathbf{n}(t^a)}{c} \times \mathbf{E}^+({\bf r},t),\;\;\;t^a = t + \frac{R(t^a)}{c},
\end{array}
\label{E_Field_ta}
\end{eqnarray}

The electric and magnetic fields of particle $j$ on particle $i$ are defined as:
\begin{eqnarray}
\begin{array}{c}
\mathbf{E}_{ji} = (1/2 + \alpha)\,\mathbf{E}^{-}_{ji} + (1/2 - \alpha)\,\mathbf{E}^{+}_{ji}, \\
\mathbf{B}_{ji} = (1/2 + \alpha)\,\mathbf{B}^{-}_{ji} + (1/2 - \alpha)\,\mathbf{B}^{+}_{ji},
\end{array}
\label{EijBij}
\end{eqnarray}
with $i\neq j$ and $\alpha\in[-1/2,1/2]$. The superscripts ``$-\,$" and ``$+\,$" indicate whether the electromagnetic field should be evaluated at retarded or advanced times respectively. The $\alpha$ parameter is just a trick to consider linear combinations between retarded and advanced fields that represent different solutions of Maxwell's equations. Two cases are worth mentioning:
the first is $\alpha = 1/2$, corresponding to the causal scenario with retarded fields;
the second is $\alpha = 0$, which corresponds to the Wheeler-Feynman formulation of electromagnetism \cite{feynman1945,feynman1949}, where  a symmetrical contribution of retarded and advanced fields is considered.

The retarded and advanced electromagnetic fields can be written as follows:
\begin{equation}
\hspace*{-12mm}
\begin{array}{cc}
\displaystyle \vspace*{2mm}\mathbf{E}^{-}_{ji} = \frac{q_{j}}{4\,\pi\,\epsilon_{0}}\,\mathbf{G}^{-}_{ji}(\mathbf{r}_{i},\mathbf{r}^{-}_{j},\mathbf{v}^{-}_{j}, \mathbf{a}^{-}_{j}),
& \displaystyle \mathbf{B}^{-}_{ji} = \frac{q_{j}}{4\,\pi\,\epsilon_{0}\,c}\,{\mathbf{n}}^{-}_{ji} 
\times \mathbf{G}^{-}_{ji}(\mathbf{r}_{i}, \mathbf{r}^{-}_{j},\mathbf{v}^{-}_{j},\mathbf{a}^{-}_{j}), \\
\displaystyle \mathbf{E}^{+}_{ji} = \frac{q_{j}}{4\,\pi\,\epsilon_{0}}\,\mathbf{G}^{+}_{ji}(\mathbf{r}_{i},\mathbf{r}^{+}_{j},\mathbf{v}^{+}_{j}, \mathbf{a}^{+}_{j}),
& \displaystyle \mathbf{B}^{+}_{ji} = \frac{q_{j}}{4\,\pi\,\epsilon_{0}\,c}\,{\mathbf{n}}^{+}_{ji} 
\times \mathbf{G}^{+}_{ji}(\mathbf{r}_{i}, \mathbf{r}^{+}_{j},\mathbf{v}^{+}_{j},\mathbf{a}^{+}_{j}),
\end{array}
\end{equation}
where
\begin{eqnarray}
\begin{array}{c}
\displaystyle\vspace*{2mm}\mathbf{G}^{-}_{ji} = \frac{(1 - \beta^{-^{2}}_{j})(\mathbf{n}^{-}_{ji} - {\beta}^{-}_{j})}{| \mathbf{r}_{i} -\mathbf{r}^{-}_{j}|^{2} |1 - \mathbf{n}^{-}_{ji} \cdot {\beta}^{-}_{j}|^{3}} + \frac{\mathbf{n}^{-}_{ji} \times [(\mathbf{n}^{-}_{ji} - {\beta}^{-}_{j}) \times \mathbf{a}^{-}_{j}]}{c^{2} |\mathbf{r}_{i} -\mathbf{r}^{-}_{j}||1 - \mathbf{n}^{-}_{ji} \cdot {\beta}^{-}_{j}|^{3}}, \\
\displaystyle\vspace*{2mm} \mathbf{G}^{+}_{ji} = \frac{(1 - \beta^{+^{2}}_{j})(\mathbf{n}^{+}_{ji} + {\beta}^{+}_{j})}{| \mathbf{r}_{i} -\mathbf{r}^{+}_{j}|^{2} |1 + \mathbf{n}^{+}_{ji} \cdot {\beta}^{+}_{j}|^{3}} + \frac{\mathbf{n}^{+}_{ji} \times [(\mathbf{n}^{+}_{ji} + {\beta}^{+}_{j}) \times \mathbf{a}^{+}_{j}]}{c^{2} |\mathbf{r}_{i} -\mathbf{r}^{+}_{j}||1 + \mathbf{n}^{+}_{ji} \cdot {\beta}^{+}_{j}|^{3}},\\
\displaystyle \mathbf{n}^{-}_{ji} = \frac{\mathbf{r}_{i} - \mathbf{r}^{-}_{j}}{\left| \mathbf{r}_{i} - \mathbf{r}^{-}_{j} \right|}, \quad \mathbf{n}^{+}_{ji} = \frac{\mathbf{r}_{i} - \mathbf{r}^{+}_{j}}{\left| \mathbf{r}_{i} - \mathbf{r}^{+}_{j} \right|},
\end{array}
\end{eqnarray}
with ${\beta}^{-}_{j}=\mathbf{v}^{-}_{j}/c$ and ${\beta}^{+}_{j}= \mathbf{v}^{+}_{j}/c$ . The dynamic variables are evaluated at their respective times:
\begin{eqnarray}
{\bf r}_i={\bf r}_i(t),\;{\bf v}_i={\bf v}_i(t),\;{\bf a}_i={\bf a}_i(t),\nonumber \\
{\bf r}^-_j={\bf r}_j(t^r_j),\;{\bf v}^-_j={\bf v}(t^r_j),\;{\bf a}^-_j={\bf a}_j(t^r_j), \label{vetores}\\
{\bf r}^+_j={\bf r}_j(t^a_j),\;{\bf v}^+_j={\bf v}(t^a_j),\;{\bf a}^+_j={\bf a}_j(t^a_j),\nonumber
\end{eqnarray}
where the retarded and advanced times satisfy the following functional equations:
\begin{equation}
t^r_j=t-\frac{|\mathbf{r}_{i}(t)-\mathbf{r}_{j}(t^r_j)|}{c}, \quad t^a_j=t+\frac{|\mathbf{r}_i(t)-\mathbf{r}_{j}(t^a_j)|}{c}.
\label{t^ret_original}
\end{equation}

Let us consider the scale transformations in space and time units defined as
\[ {\bf r}_i\rightarrow L{\bf r}_i\,,\;\; t\rightarrow Tt,\]
where $L$ and $T$ are given by:
\[  L={\frac { \left| { q_1} \right|  \left| { q_2} \right| }{4\pi \epsilon_0{ m_2}{c}^{2}\\
\mbox{}}},\,\, T={\frac { \left| { q_1} \right|  \left| { q_2} \right| }{4\pi \epsilon_0{ m_2}{c}^{3}}}.\]
This scale transformation leads to the following transformations in velocities, moments, and accelerations:
\[ {\bf v}_i\rightarrow c{\bf v}_i,\;\; {\bf p}_i\rightarrow m_2 c \,{\bf p}_i,\;\;{\bf a}_i\rightarrow \frac{c}{T}{\bf a}_i.  \]
It is worth emphasizing that  $L/T=c$ and  $c=1$ in the new units of length and time.

The scale transformations applied to Eq. (\ref {newton_tr_ta}) leads to
\begin{eqnarray}
\label{eqG1}
\hspace*{-13mm}\frac{d{\bf p}_1}{dt}=\eta\,\frac{d{(\gamma_{1}\,\mathbf{v}_{1})}}{dt} = S\,[(1/2+\alpha)\,\mathbf{G}^{-}_{21}+(1/2-\alpha)\,\mathbf{G}^{+}_{21}] \nonumber \\
\hspace*{23mm}+S\,\mathbf{v}_{1} \times [(1/2+\alpha)\,\mathbf{n}^{-}_{21} \times \mathbf{G}^{-}_{21} + (1/2-\alpha)\,\mathbf{n}^{+}_{21} \times\mathbf{G}^{+}_{21}], \\
\hspace*{-13mm}\frac{d{\bf p}_2}{dt}=\frac{d{(\gamma_{2}\,\mathbf{v}_{2})}}{dt} = S\,[(1/2+\alpha)\,\mathbf{G}^{-}_{12} + (1/2-\alpha)\,\mathbf{G}^{+}_{12}] \nonumber \\ 
\hspace*{23mm}+ S\,\mathbf{v}_{2} \times [(1/2+\alpha)\,\mathbf{n}^{-}_{12} \times \mathbf{G}^{-}_{12} + (1/2-\alpha)\,\mathbf{n}^{+}_{12} \times\mathbf{G}^{+}_{12}], \nonumber
\end{eqnarray}
where
$${\bf p}_i=\eta\gamma_1{\bf v}_i,\; \gamma_{i}= (1-|{\mathbf{v}}_{i}|^2)^{-1/2}\; ,$$
$$\eta=m_1/m_2,\; S=\mbox{\rm sgn}(q_1q_2)=\pm 1.$$
We should consider $c=1$ in the formulas for ${\bf G}^{\, -}_{ij}$,  ${\bf G}^{\, +}_{ij}$ and the functional equations for retarded and advanced times.
We note that the system of equations (\ref{eqG1}) is dimensionless and depends only on three parameters: $\eta$, $S$ and 
$\alpha$.

\subsection{The Explicit Second Order Newton Equations as a Neutral System of FDEs} 

The Eq. (\ref{eqG1}) can be written in a matrix form as the following second-order dynamical system:
\begin{eqnarray}
\hspace*{-13mm}\frac{d{\bf p}_1}{dt}=\eta \mathcal{M}_{11} \mathbf{a}_{1} = (1/2+\alpha)\left(\mathbf{F}^{-}_{1} - \mathcal{M}^{-}_{12} \mathbf{a}^{-}_{2}\right)
+ (1/2-\alpha)\left(\mathbf{F}^{+}_{1} - \mathcal{M}^{+}_{12} \mathbf{a}^{+}_{2}\right)  \nonumber\\
\hspace*{-13mm}\frac{d{\bf p}_2}{dt}= \mathcal{M}_{22} \mathbf{a}_{2} =  (1/2+\alpha)\left(\mathbf{F}^{-}_{2} - \mathcal{M}^{-}_{21} \mathbf{a}^{-}_{1}\right)
+ (1/2-\alpha)\left(\mathbf{F}^{+}_{2} - \mathcal{M}^{+}_{21} \mathbf{a}^{+}_{1}\right).  
\label{eq2_alpha_pontual}
\end{eqnarray}
The vector ${\bf F}^-_j$ ($j=1,2$) and the matrices ${\cal M}^-_{ij}$ ($i,j=1,2$ with $j\neq i$) depends on ${\bf r}_i(t)$,   ${\bf r}_j(t^r_j)$, ${\bf v}_i(t)$ and ${\bf v}_j(t^r_j)$.  
The vector ${\bf F}^+_j$ ($j=1,2$) and the matrices ${\cal M}^+_{ij}$ ($i,j=1,2$ with $j\neq i$) depends on ${\bf r}_i(t)$,   ${\bf r}_j(t^a_j)$, ${\bf v}_i(t)$ and ${\bf v}_j(t^a_j)$.
The explicit expressions for these vectors and matrices are given respectively in Appendix A (retarded fields) and Appendix B (advanced fields).  
The matrices  $\mathcal{M}_{ii}$ ($i=1,2$) are evaluated at time $t$ and defined as follows:  
\begin{eqnarray}
\mathcal{M}_{ii} = \gamma_{i}\mathcal{I} + \gamma^{3}_{i}|{\bf v}_i|^2 {\cal Q}_{\hat{\bf v}_{i}}\;\;(i=1,2), 
\end{eqnarray}
where $\hat{\bf v}_{i} = \mathbf{v}_{i} / |\mathbf{v}_{i}|$. The action of the matrix  ${\cal Q}_{\hat{\bf w}}$ (called projector operator) in any vector ${\bf f}$ is defined as ${\cal Q}_{\hat {\bf w}}{\bf f}=({\bf f}\cdot{\hat{\bf w}}){\hat{\bf w}}$.
The matrix $\mathcal{I}$ stand for the $3$-dimensional identity matrix.

The self-force ${\bf F}^{self}$ of the two point charges system  is defined as the time derivation of its total moment, which from Eq. (\ref{eq2_alpha_pontual}) is given by:
\begin{eqnarray}
\hspace*{-15mm}{\bf F}^{self}=\frac{d\left({\bf p}_1+{\bf p}_2\right)}{dt}=
\left(1/2+\alpha\right)\left[\mathbf{F}^{-}_{1} +\mathbf{F}^{-}_{2}-\left( \mathcal{M}^{-}_{21} \mathbf{a}_1(t_1^r)+\mathcal{M}^{-}_{12} \mathbf{a}_2(t_2^r)\right)\right]\nonumber\\
\hspace*{26mm}+\left(1/2-\alpha\right)\left[\mathbf{F}^{+}_{1} +\mathbf{F}^{+}_{2}-\left( \mathcal{M}^{+}_{21} \mathbf{a}_1(t_1^a)+\mathcal{M}^{+}_{12} \mathbf{a}_2(t_2^a)\right)\right].
\label{auto_forca}
\end{eqnarray}
The self-force is the resultant force due exclusively to the electromagnetic fields associated to the interaction between the charges. Differently from the traditional classical Newtonian mechanics, the third Newton's law is not valid and  the total mechanical moment is not conserved.

If we inverse the matrices $\mathcal{M}_{11}$ and $\mathcal{M}_{22}$, then  Eq. (\ref{eq2_alpha_pontual})  becomes an explicitly second-order dynamical system:
\begin{eqnarray}
\label{a1-delay-antigo}
\hspace*{-15mm}\eta \,\mathbf{a}_{1}=(1/2+\alpha) \mathcal{M}^{-1}_{11}
\left[\mathbf{F}^{-}_{1} - \mathcal{M}^{-}_{12}\mathbf{a}_{2}(t_2^r)\right]+(1/2-\alpha)\mathcal{M}^{-1}_{11}
\left[\mathbf{F}^{+}_{1} - \mathcal{M}^{+}_{12}\mathbf{a}_{2}(t_2^a)\right]\nonumber\\
\hspace*{-15mm}\mathbf{a}_{2}= (1/2+\alpha)\mathcal{M}^{-1}_{22}
\left[\mathbf{F}^{-}_{2} - \mathcal{M}^{-}_{21}\mathbf{a}_{1}(t_1^r)\right]+ (1/2-\alpha)\mathcal{M}^{-1}_{22}
\left[\mathbf{F}^{+}_{2} - \mathcal{M}^{+}_{21}\mathbf{a}_{1}(t_1^a)\right].
\end{eqnarray}
The matrices ${\cal Q}_{\hat{\bf v}_{1}}$ and ${\cal Q}_{\hat{\bf v}_{2}}$ are idempotents and allows us to show that
\begin{equation}
\label{matrizesMii}
\mathcal{M}^{-1}_{ii}(t) = \left(1- \left| \mathbf{v}_{i}(t) \right|^{2}\right)^{1/2}
\left( \mathcal{I} - \left| \mathbf{v}_{i}(t) \right|^{2}{\cal Q}_{\hat{\bf v}_{i}}(t)\right),\; i=1,2.
\end{equation}
The respective retarded and advanced times obey the following functional equations: 
\begin{eqnarray}
t^r_1=t-|\mathbf{r}_{2}(t)-\mathbf{r}_{1}(t^r_1)|, &\quad& t^a_1=t+|\mathbf{r}_2(t)-\mathbf{r}_{1}(t^a_1)|,\nonumber\\
t^r_2=t-|\mathbf{r}_{1}(t)-\mathbf{r}_{2}(t^r_2)|, &\quad& t^a_2=t+|\mathbf{r}_1(t)-\mathbf{r}_{2}(t^a_2)|.
\label{t^ret^rescaled}
\end{eqnarray}

The Eqs. (\ref{a1-delay-antigo}) and (\ref{t^ret^rescaled}) constitute a neutral system of  FDEs, where the four functional equations  depend on the particle trajectories. 
The respective system of FDEs can be written as:
\begin{eqnarray}
\label{a1-delay}
\eta{\bf a}_1(t)&=&{\bf W}_1\left[{\bf r}_1(t),{\bf v}_1(t),{\bf r}_2(t_2^r),{\bf v}_2(t_2^r),{\bf a}_2(t_2^r),{\bf r}_2(t_2^a),{\bf v}_2(t_2^a),{\bf a}_2(t_2^a)\right]\nonumber\\  
{\bf a}_2(t)&=&{\bf W}_2\left[{\bf r}_2(t),{\bf v}_2(t),{\bf r}_1(t_1^r),{\bf v}_1(t_1^r),{\bf a}_1(t_1^r),{\bf r}_1(t_1^a),{\bf v}_1(t_1^a),{\bf a}_1(t_1^a)\right]\\
&&\hspace*{-5mm}t^r_1=t-|\mathbf{r}_{2}(t)-\mathbf{r}_{1}(t^r_1)|, \quad t^a_1=t+|\mathbf{r}_2(t)-\mathbf{r}_{1}(t^a_1)|,\nonumber\\
&&\hspace*{-5mm}t^r_2=t-|\mathbf{r}_{1}(t)-\mathbf{r}_{2}(t^r_2)|, \quad t^a_2=t+|\mathbf{r}_1(t)-\mathbf{r}_{2}(t^a_2)|,\nonumber
\end{eqnarray}
where the vector fields ${\bf W}_i$ ($i=1,2$) are directly obtained from Eq. (\ref{a1-delay-antigo}).
The Synge's problem consists on the search for trajectories that satisfy this neutral system. Indeed, this problem concerns to the existence of global solutions. 
Let us remark to the crossed dependency of accelerations at present time with respect to the variables at retarded and advanced times, that is, the acceleration of one particle at present time depends on the variables of the other particle at retarded and advanced times. We explicit write the dependence of variables on present time $t$ for sake of clarity.

\section{A new iterative method to solve Synge's problem}

Our goal is to develop a method to built a sequence of ODEs system in order to obtain sucessive approximate solutions for the FDEs system  in Eq. (\ref{a1-delay}).  To simplify the method description, we present the FDEs system (\ref{a1-delay}) in a more compact form. For this purpose we
define the state vector ${\bf X}=\left({\bf r}_1,{\bf v}_1,{\bf r}_2,{\bf v}_2\right)$ and the vector of extended state $\bar{\bf X}=\left({\bf r}_1,{\bf v}_1,{\bf a}_1,{\bf r}_2,{\bf v}_2,{ \bf a}_2\right)$, then the FDEs system (\ref{a1-delay}) can be rewritten as a first-order system with the following general form:
\begin{equation}
\begin{array}{c}
\displaystyle \frac{d {\bf X}(t)}{dt}={\bf G}\left[{\bf X}(t),\bar{\bf X}(t_1^r),\bar{\bf X}(t_2^r),\bar{\bf X}(t_1^a),\bar{\bf X}(t_2^a)\right], \\
t_1^r=t-|\mathbf{r}_{2}(t)-\mathbf{r}_{1}(t_1^r)|, \quad t_2^r=t-|\mathbf{r}_1(t)-\mathbf{r}_{2}(t_2^r)|, \\
t_1^a=t+|\mathbf{r}_{2}(t)-\mathbf{r}_{1}(t_1^a)|, \quad t_2^a=t+|\mathbf{r}_1(t)-\mathbf{r}_{2}(t^a_2)|,
\end{array}
\label{eqWret}
\end{equation}
where the vector field ${\bf G}$ is uniquely obtained from equation (\ref{a1-delay}).

Considering $t_1^r=t_2^r=t^a_1= t^a_2=t$ in equation (\ref{a1-delay}), we obtain a linear system for the accelerations ${\bf a}_i$ ($i=1, 2$). This system can be  solved leading to a explicit system of second-order ODEs that, in turn, can be transformed into a system of first-order ODEs. We write this first-order system compactly as
\begin{equation}
\label{eqH0}
\frac{d{\bf X}}{dt}={\bf H}^{(0)}\left({\bf X}\right).
\end{equation}
The system (\ref{eqH0}) represents what we call  the instantaneous approximation, that is, all variables are calculated at the current time $t$.

We define the extended flow 
\begin{equation}
\phi^{(0)}_{{\bf X}}(\tau)=\left({\bf r}^{(0)}_1(\tau),{\bf v}^{(0)}_1(\tau),{\bf a}^{(0)}_1(\tau),{\bf r}^{(0)}_2(\tau),{\bf v}^{(0)}_2(\tau),{\bf a}^{(0)}_2(\tau)\right),
\end{equation}
which stands for the unique maximal solution of the system (\ref{eqH0}) with initial condition  $\phi^{(0)}_{\bf X}(0)={\bf X}$. The time $\tau$ is defined for the maximal interval where the maximal solution is determined.
Thus, we get an approximation for the retarded and advanced times in Eq. (\ref{eqWret})  by solving the following equations:
\begin{eqnarray}
\tau^{(0)}_1=t-|\mathbf{r}_{2}(t)-\mathbf{r}^{(0)}_{1}(\tau^{(0)}_1)|, &\quad& \tau^{(0)}_2=t-|\mathbf{r}_1(t)-\mathbf{r}^{(0)}_{2}(\tau^{(0)}_2)|,\nonumber\\
\bar\tau^{(0)}_1=t+|\mathbf{r}_{2}(t)-\mathbf{r}^{(0)}_{1}(\bar\tau^{(0)}_1)|, &\quad& \bar\tau^{(0)}_2=t+|\mathbf{r}_1(t)-\mathbf{r}^{(0)}_{2}(\bar\tau^{(0)}_2)|,
\end{eqnarray}
builted from the extended flow $\phi^{(0)}_{{\bf X}(t)}(\tau)$.
Replacing
\begin{eqnarray}
\bar{\bf X}(t_1^r)=\phi^{(0)}_{{\bf X}(t)}(\tau^{(0)}_1),&\;\;&\bar{\bf X}(t_2^r)=\phi^{(0)}_{{\bf X}(t)}(\tau^{(0)}_2),\nonumber \\
\bar{\bf X}(t_1^a)=\phi^{(0)}_{{\bf X}(t)}(\bar\tau^{(0)}_1),&\;\;&\bar{\bf X}(t_2^a)=\phi^{(0)}_{{\bf X}(t)}(\bar\tau^{(0)}_2),
\end{eqnarray}
into equation (\ref{eqWret}) we get the system of ODEs given by
\begin{equation}
\frac{d {\bf X}(t)}{dt}={\bf G}\left[{\bf X}(t),\phi^{(0)}_{{\bf X}(t)}(\tau^{(0)}_1),\phi^{(0)}_{{\bf X}(t)}(\tau^{(0)}_2),
\phi^{(0)}_{{\bf X}(t)}(\bar\tau^{(0)}_1),\phi^{(0)}_{{\bf X}(t)}(\bar\tau^{(0)}_2)\right].
\label{eqWauto}
\end{equation}

We observe that $\tau^{(0)}_1$, $\tau^{(0)}_2$, $\bar\tau^{(0)}_1$ and $\bar\tau^{(0) }_2$ are  uniquely determined by the state vector ${ \bf X}(t)$ evaluated at the current time $t$. Therefore, we remove the explicitness of variables that depend on the current time and rewrite, without loss of generality, the system (\ref{eqWauto}) as
\begin{eqnarray}
&\displaystyle \frac{d {\bf X}}{dt}={\bf H}^{(1)}({\bf X})={\bf G}\left[{\bf X},\phi^{(0)}_{{\bf X}}(\tau^{(0)}_1),\phi^{(0)}_{{\bf X}}(\tau^{(0)}_2),
\phi^{(0)}_{{\bf X}}(\bar\tau^{(0)}_1),\phi^{(0)}_{{\bf X}}(\bar\tau^{(0)}_2)\right],& \label{eqH1}\nonumber \\
&\tau^{(0)}_1=-|\mathbf{r}_{2}-\mathbf{r}^{(0)}_{1}(\tau^{(0)}_1)|, \quad \tau^{(0)}_2=-|\mathbf{r}_1-\mathbf{r}^{(0)}_{2}(\tau^{(0)}_2)|.& \\
&\bar\tau^{(0)}_1=|\mathbf{r}_{2}-\mathbf{r}^{(0)}_{1}(\bar\tau^{(0)}_1)|, \quad \bar\tau^{(0)}_2=|\mathbf{r}_1-\mathbf{r}^{(0)}_{2}(\bar\tau^{(0)}_2)|.& \nonumber
\end{eqnarray}
The ODEs system (\ref{eqH1}) can be seen as an approximation of the FDEs system in (\ref{eqWret}). 

Indeed, the procedure above has used the ODEs system
(\ref{eqH0}) to obtain approximations for  variables calculated  at retarded and advanced times in the FDEs system (\ref{eqWret}), allowing to define a new ODEs system in Eq.  (\ref{eqH1}).
If we think about this procedure as transforming a system of ODEs into another system of ODEs, we could apply it analogously to the system (\ref{eqH1}), which would  play the role of the system (\ref {eqH0}), to obtain a new system of ODEs from Eq. (\ref{eqWret}).

Formally, we can think this procedure as a ${\cal T}$ map that generates a sequence of vector fields  (ODEs systems, respectively):
\begin{equation}
\frac{d{\bf X}}{dt}={\bf H}^n({\bf X})\;(n=0,1,2,\ldots)\;\;\Rightarrow\;\;{\bf H}^{(n)}({\bf X})\stackrel{\displaystyle \cal T}{\longrightarrow} {\bf H}^{(n+1)}({\bf X})
\label{sequenciaHn}
\end{equation}

The sequence of  ODEs systems in (\ref{sequenciaHn}) is built tacking into account the following conditions.
\begin{itemize}
\item For a given compact set $U$, there is a compact set $U^{(n)}\supset U$ such that the vector field ${\bf H}^{(n)}({\bf X})$ is  differentiable for all ${\bf X}\in {U}^{(n)}$.
\item For all ${\bf X}\in U$ there are real numbers: $\tau_1^{(n)}$, $\bar\tau_1^{(n)}$, $\tau_2^{(n)}$ and $\bar\tau_2^{(n)}$  satisfying the equations
\begin{eqnarray}
&\tau^{(n)}_1=-|\mathbf{r}_{2}-\mathbf{r}^{(n)}_{1}(\tau^{(n)}_1)|, \quad \tau^{(n)}_2=-|\mathbf{r}_1-\mathbf{r}^{(n)}_{2}(\tau^{(n)}_2)|,& \nonumber\\
&\bar\tau^{(n)}_1=|\mathbf{r}_{2}-\mathbf{r}^{(n)}_{1}(\bar\tau^{(n)}_1)|, \quad \bar\tau^{(n)}_2=|\mathbf{r}_1-\mathbf{r}^{(n)}_{2}(\bar\tau^{(n)}_2)|,& 
\label{temp_aproximado}
\end{eqnarray}
where 
\begin{equation}
\label{fluxo-n}
\phi^{(n)}_{{\bf X}}(\tau)=\left({\bf r}^{(n)}_1(\tau),{\bf v}^{(n)}_1(\tau),{\bf a}^{(n)}_1(\tau),{\bf r}^{(n)}_2(\tau),{\bf v}^{(n)}_2(\tau),{\bf a}^{(n)}_2(\tau)\right)
\end{equation}
is the flow of solutions associated to the ODEs system defined by the vector field ${\bf H}_n({\bf X})$ with ${\bf X}\in U$, and
\begin{equation}
\label{variavel-fluxo-n}
\phi^{(n)}_{{\bf X}}(\tau^{(n)}_1),\, \phi^{(n)}_{{\bf X}}(\tau^{(n)}_2),\,\phi^{(n)}_{{\bf X}}(\bar\tau^{(n)}_1),\,\phi^{(n)}_{{\bf X}}(\bar\tau^{(n)}_2)\,\in U^{(n)}.
\end{equation} 
\item The sequence of vector fields in (\ref{sequenciaHn}) is initialized with the instantaneous vector field ${\bf H}^{(0)}(\bf X)$ given in Eq. (\ref{eqH0}).
\end{itemize}
Then, we can define the ODEs system
\begin{eqnarray}
\label{sistemaHn}
\hspace*{-2mm}\displaystyle \frac{d {\bf X}}{dt}={\bf H}^{(n+1)}({\bf X})={\bf G}\left[{\bf X},\phi^{(n)}_{{\bf X}}(\tau^{(n)}_1),\phi^{(n)}_{{\bf X}}(\tau^{(n)}_2),
\phi^{(n)}_{{\bf X}}(\bar\tau^{(n)}_1),\phi^{(n)}_{{\bf X}}(\bar\tau^{(n)}_2)\right]
\end{eqnarray}
for all ${\bf X}\in U$, where ${\bf G}$ is given in Eq.  (\ref{eqWret}). 
 
Let us suppose that the sequence of vector fields defined in (\ref{sequenciaHn}) converges for all ${\bf X}\in U$, that is,
\begin{equation}
\lim_{n\rightarrow\infty}{\bf H}^n({\bf X})={\bf H}^{\infty}({\bf X})\;\;\;\forall\,{\bf X}\in U.
\label{convergenciaH}
\end{equation}
Thus,
\begin{equation}
\phi^{(\infty)}_{\bf X}(\tau)=\lim_{n\rightarrow\infty}\phi^{(n)}_{\bf X}(\tau)
\label{convergenciaFluxo}
\end{equation}
is the extended flow of solutions for the ODEs system defined by the vector field ${\bf H}^{\infty}({\bf X})$ on $U$.
Indeed, the flow of solutions in Eq. (\ref{convergenciaFluxo}) defines a set of  solutions for the FDEs system in (\ref{a1-delay}) with initial conditions belonging to  $U$. In this case,  
the solutions for the ODEs systems in the sequence (\ref{sequenciaHn})
represent successive approximations to global Synge solutions that intercept the set $U$.

\subsection{The Instantaneous System defined by the vector field ${\bf H}^{(0)}({\bf X})$}

The system of ODEs defined in (\ref{eqH0}), called instantaneous system, can be explicitly obtained considering that all variable are calculated at the current time $t$, that is,
we consider the approximation $t_1^r=t_2^r=t_1^a=t_2^a=t$ in the system of differential equations defined in (\ref{a1-delay-antigo}). Therefore, this system can be explicitly written as
\begin{eqnarray}
\hspace*{-22mm}\eta \,\mathbf{a}_{1}&=&(1/2+\alpha) \mathcal{M}^{-1}_{11}
\left[\mathbf{F}^{-}_{1}(t) - \mathcal{M}^{-}_{12}(t)\mathbf{a}_{2}\right]
+(1/2-\alpha)\mathcal{M}^{-1}_{11}
\left[\mathbf{F}^{+}_{1}(t) - \mathcal{M}^{+}_{12}(t)\mathbf{a}_{2}\right],\nonumber\\
\hspace*{-22mm}\mathbf{a}_{2}&=& (1/2+\alpha)\mathcal{M}^{-1}_{22}
\left[\mathbf{F}^{-}_{2}(t) - \mathcal{M}^{-}_{21}(t) \mathbf{a}_{1}\right] 
+ (1/2-\alpha)\mathcal{M}^{-1}_{22}
\left[\mathbf{F}^{+}_{2}(t) - \mathcal{M}^{+}_{21}(t) \mathbf{a}_{1}\right],
\label{sist_inst_implicito}
\end{eqnarray}
where $\mathbf{F}^{-}_{i}(t)$, $\mathbf{F}^{+}_{i}(t)$, $\mathcal{M}^{ -}_{ij}(t)$ and $\mathcal{M}^{+}_{ij}(t)$ ($i,j=1,2\;j\neq i$) can be explicitly calculated from their expressions defined  in Appendices A and B.

The system in Eq. (\ref{sist_inst_implicito}) is an implicit second-order ODEs system,  defined through a system of linear equations for the particle  accelerations.
We can solve this system, obtaining the acceleration as functions of positions and velocities of particles. After some straightforward algebra, we can show that
\begin{eqnarray}
\hspace*{-15mm}\eta \mathbf{a}_{1} &=& \left( \mathcal{I} - \frac{1}{\eta} \mathcal{M}^{-1}_{11}\mathcal{M}^\alpha_{12} \mathcal{M}^{-1}_{22} \mathcal{M}^\alpha_{21} \right)^{-1} \left(   \mathcal{M}^{-1}_{11} \mathbf{F}^\alpha_{1} - \mathcal{M}^{-1}_{11} \mathcal{M}^\alpha_{12} \mathcal{M}^{-1}_{22} \mathbf{F}^\alpha_{2}  \right), \nonumber\\
\hspace*{-15mm}\mathbf{a}_{2} &=& \left( \mathcal{I} - \frac{1}{\eta} \mathcal{M}^{-1}_{22}\mathcal{M}^\alpha_{21} \mathcal{M}^{-1}_{11} \mathcal{M}^\alpha_{12} \right)^{-1} \left(  \mathcal{M}^{-1}_{22} \mathbf{F}^\alpha_{2} - \frac{1}{\eta}\mathcal{M}^{-1}_{22} \mathcal{M}^\alpha_{21} \mathcal{M}^{-1}_{11} \mathbf{F}^\alpha_{1}  \right),
\label{sist_inst_explicito} 
\end{eqnarray}
where ${\bf F}^\alpha_i$ and ${\cal M}^\alpha_{ij}$ have expressions given by
\begin{eqnarray}
{\bf  F}^\alpha_i&=&\left(\frac{1}{2}+\alpha\right){\bf F}^-_i(t)+\left(\frac{1}{2}-\alpha\right){\bf F}^+_i(t)\;\;(i=1,2),
\label{expressoesF}\\
{\cal  M}^\alpha_{ij}&=&\left(\frac{1}{2}+\alpha\right){\cal M}^-_{ij}(t)+\left(\frac{1}{2}-\alpha\right){\cal M}^+_{ij}(t)\;\;(i,j=1,2\; j\neq i).
\label{expressoesM}
\end{eqnarray}
The vectors and matrices respectively defined in (\ref{expressoesF}) and (\ref{expressoesM}) depend only on the positions and velocities of the particles, having no explicit dependence on the current time $ t$. Therefore, the ODEs system defined in (\ref{sist_inst_explicito}) is also an autonomous system.

From the $6$-dimensional system of second-order ODEs in (\ref{sist_inst_explicito}) we  obtain the vector field ${\bf H}^{(0)}$, which defines a $12$-dimensional system of first order in (\ref{eqH0}). This system allows to calculate the flow of solutions $\phi^{(0)}_{{\bf X}}(\tau)$, and then to obtain the  entire sequence of vector fields 
in (\ref{sequenciaHn}).

\subsection{A numerical method to integrate the system defined by the vector field ${\bf H}^{(n+1)}({\bf X})$}

The problem to build a numerical integration method for the system of ODEs  in the sequence (\ref{sequenciaHn}) is to obtain well-defined expression for the vector field ${\bf H}^{(n+1)}( {\bf X})$  for a given state vector ${\bf X}=({\bf r}_1,{\bf v}_1,{\bf r}_2,{\bf v}_2)$. Let us  assume that the extended flow in Eq. (\ref{fluxo-n})
is known, that is,  we are assuming that this extended flow can be numerically calculated for each vector ${\bf X}$ and time $\tau$ considered.

Therefore, the first task  is to solve the functional equations  in (\ref{temp_aproximado}), provided we can calculate $\mathbf{r}^{(n)}_{1}(\tau)$ and $\mathbf{r}^{(n)}_{2}(\tau)$ from the extended flow  in (\ref{fluxo-n}). 
To obtain the values of $\tau^{(n)}_1$, $\tau^{(n)}_2$, $\bar\tau^{(n)}_1$ and $\bar\tau^{(n)}_2$, we solve respectively the four functional equations in (\ref{temp_aproximado}) using the bisection algorithm \cite{kalu2002}. Furthemore, we calculate the vector field ${\bf H}^{(n+1)}({\bf X})$ in (\ref {sistemaHn}) from the positions, velocities and accelerations obtained through the extended flow calculated at these times, according to Eq. (\ref{variavel-fluxo-n}).

Once built the algorithm for calculating the vector field ${\bf H}^{(n+1)}({\bf X})$,  we  implement the Runge-Kutta method of order $4$-$5$, developed by Fehlberg \cite{fehlberg1969,shampine1976}, to calculate the extended flow $\phi^{(n+1)}_{{\bf X}}(\tau)$. The procedure is established recursively: for a given state
${\bf X}$, the field ${\bf H}^{(n+1)}({\bf X})$ is obtained from the extended flow $\phi^{(n)} _{{\bf X}}(\tau)$, this in turn depends on the calculation of the field ${\bf H}^{(n)}({\bf X})$, which in turn depends on the
extended flow $\phi^{(n-1)}_{{\bf X}}(\tau)$, and so on recursively until we have to initially compute the extended flow $\phi^{(0)}_{{\bf X}}(\tau)$, implemented to calculate the solutions of the ODEs system defined by the field ${\bf H}^{(0)}({\bf X})$.
  
To obtain the vector field ${\bf H}^{(n+1)}({\bf X})$,  the extended flux (\ref{fluxo-n}) must be calculated at the respective times that solve the functional equations in (\ref{temp_aproximado}). That happens because this vector field depends on the particle accelerations at  retarded and advanced times, as one can clearly see in  Eq.  (\ref{a1-delay}).  Therefore, once the flow of solutions has been calculated for a given state ${\bf X}$ and  time $\tau$, we need to evaluate the accelerations at this time in order to obtain the extended flow (\ref{fluxo-n}).

The calculation of  accelerations is done approximately as follows: given a state ${\bf X}$ and time $\tau$, the Runge-Kutta algorithm obtains the particle velocities at the time $\tau$ for the solution with initial condition given by ${\bf X}$; 
we define a small value $\Delta \tau$ ($\Delta\tau<<\tau$) in order to calculate the velocities
at the time $\tau+\Delta\tau$, with the same solution corresponding to the initial state ${\bf X}$;  finally, we estimate an approximate value for the particle accelerations as being
$${\bf a}_i(\tau)=\frac{{\bf v}_i(\tau+\Delta\tau)-{\bf v}_i(\tau)}{\Delta\tau}.$$

All numerical algorithms  were implemented in C++  and, in the next section, we apply them to study some planar solutions for the Synge Problem. 

\section{Application to bidimensional (planar) system}

We apply the method described in the  previous section to find numerical solutions for planar systems, that is, trajectories  in a given Cartesian coordinates $(x,y,z)$ and restricted to the plane $z=0$. If  the $z$-component  of positions, velocities and accelerations is null  in the right side of equation (\ref{a1-delay}), then the component $z$ of accelerations, in the left side of this equation, is also null. Therefore, equation  (\ref{a1-delay})  becomes a planar system of differential equations, constituting a system of four second order equations defined in the bidimensional cartesian plane $(x,y)$.  The respective formulas for the forces that define this planar system is given in Appendix C.  

We will show trajectories for attractive systems ($S=-1$) with retarded electromagnetic fields ($\alpha=1/2$) and symmetric electromagnetic fields ($\alpha=0$).
For all trajectories, we consider initial conditions such that the distance between particles is $||{\bf r}||=50$ and that would imply circular trajectories, since only the non-relativistic instantaneous Coulomb force is considered.

\subsection{Retarded Eletromanetic Fileds ($\alpha=1/2$)}

For retarded and attractive fields, the particles tend to get closer until they collide after a certain finite time (called singularity time), and the velocity of the lighter particle tend to increase up to the light speed \cite{ synge1940}.  The numerical integrations are carried out up to the velocity of the fastest particle attain $80\%$ of the light speed.  For each $n$ in the sequence of ODEs system given in Eq. (\ref{sequenciaHn}), we denote the total integration time as  $t^{(n)}$.

\begin{figure}
[!htb]
\begin{center}
\hspace*{-10mm}\includegraphics[width= 17.0cm]{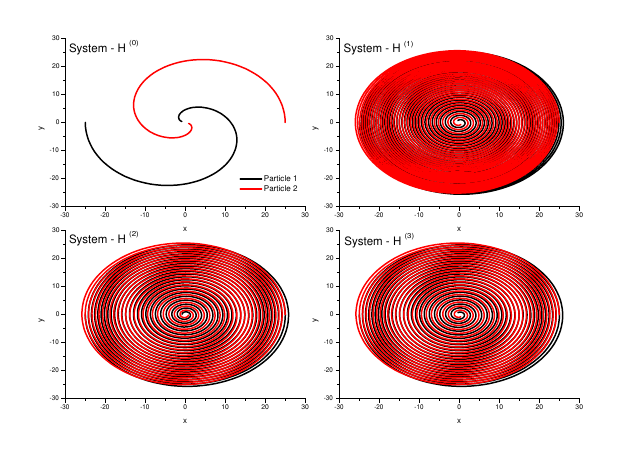}
\end{center}
\vspace*{-15mm}
\caption{\footnotesize Planar trajectories in Cartesian coordinates $(x,y)$ for the vectors ${\bf r}_1$, ${\bf r}_2$, where we have integrate numerically (considering the same initial condition) the first four ODEs systems of the sequence (\ref{sequenciaHn}) with $\eta=1$. Also, we can observe that the particle trajectories of the ${\bf H}^{(2)}$ system are practically identical to that of the ${\bf H}^{(3)}$ system}
\label{pai1}
\end{figure}

\begin{figure}
[!htb]
\begin{center}
\hspace*{-10mm}\includegraphics[width= 17.0cm]{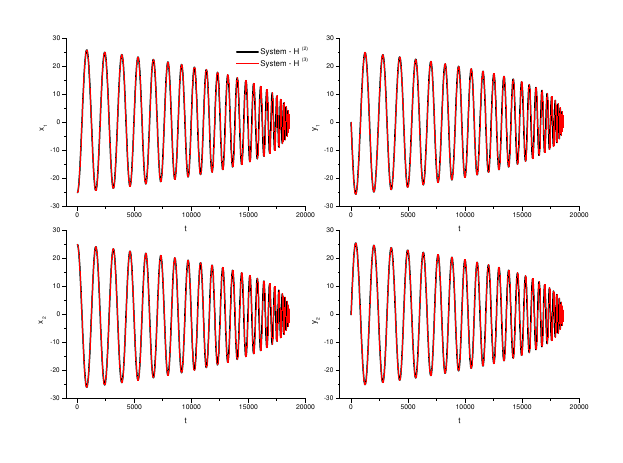}
\end{center}
\vspace*{-15mm}
\caption{\footnotesize Comparison of  particle position components in a Cartesian system $(x,y)$ for trajectories integrated from systems  ${\bf H}^{(2)}$ and  ${\bf H}^{(3)}$ for $\eta=1$. The trajectories here are those shown in the respective panels System ${\bf H}^{(2)}$ and System ${\bf H}^{(3)}$ of figure \ref{pai1}.  Let us remember that ${\bf r}_1=(x_1,y_1)$ and ${\bf r}_2=(x_2,y_2)$.}
\label{pai2}
\end{figure}

Figure \ref{pai1} shows trajectories for systems of two charges with equal masses $(\eta=1)$, corresponding to the first four systems in the sequence (\ref{sequenciaHn}), that is,  we are considering $n=0,1,2,3$.
We can see that the trajectories, for the systems defined by the vector fields ${\bf H}^{(2)}$ and ${\bf H}^{(3)}$, are very similar. Indeed, the trajectories convergence appears to be quite fast for the quasi-circular spiral trajectory considered.  This convergence can be better observed in figure \ref{pai2}, where we compare the particle trajectories obtained from the systems of ODEs defined respectively by ${\bf H}^{(2)}$ and ${\bf H}^{(3) }$.

\begin{figure}
[!htb]
\begin{center}
\hspace*{-10mm}\includegraphics[width= 17.0cm]{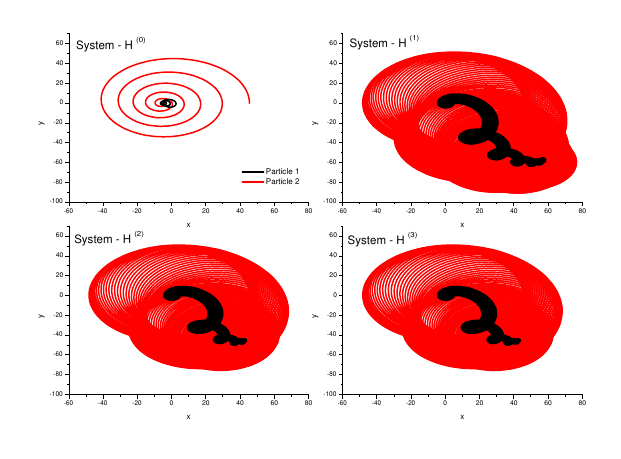}
\end{center}
\vspace*{-15mm}
\caption{\footnotesize Planar trajectories in Cartesian coordinates $(x,y)$ for the vectors ${\bf r}_1$, ${\bf r}_2$, where we have integrate numerically (for the same initial condition) the first four ODEs systems of the sequence (\ref{sequenciaHn}) considering  $\alpha=1/2$ and 
$\eta=10$. We can observe that the particle trajectories of the ${\bf H}^{(2)}$ system are practically identical to that of the ${\bf H}^{(3)}$ system}
\label{pai3}
\end{figure}

\begin{figure}
[!htb]
\begin{center}
\hspace*{-10mm}\includegraphics[width= 17.0cm]{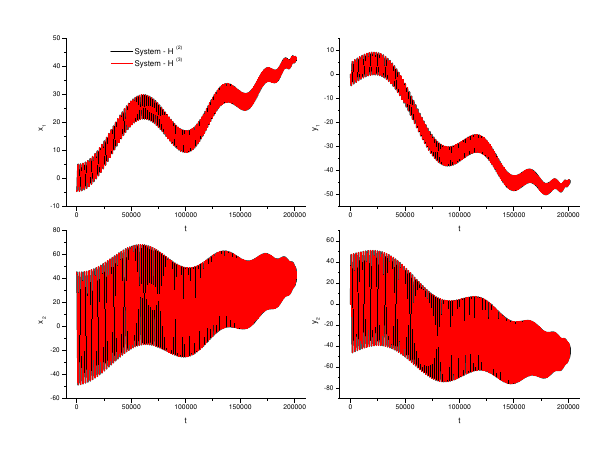}
\end{center}
\vspace*{-15mm}
\caption{\footnotesize Comparison of  particle position components in a Cartesian system $(x,y)$ for trajectories integrated from systems  ${\bf H}^{(2)}$ and  ${\bf H}^{(3)}$ for $\eta=10$. The trajectories here are the same as those shown in the respective panels System ${\bf H}^{(2)}$ and 
System ${\bf H}^{(3)}$ of figure \ref{pai3}.  Let us remember that ${\bf r}_1=(x_1,y_1)$ and ${\bf r}_2=(x_2,y_2)$.}
\label{pai4}
\end{figure}

Figure \ref{pai3} shows trajectories for $\eta=10$.
As in the case for $\eta=1$, we can see that the trajectories of systems ${\bf H}^{(2)}$ and ${\bf H}^{(3)}$ are almost the same.
This fact can be observed in figure \ref{pai4}, where we compare the evolution for particle trajectories of systems ${\bf H}^{(2)}$ and ${\bf H}^{(3)}$.
We can observe in figure \ref{pai3} the movement of the centre of mass: we see that particles, respectively the center of mass, tend to move globally to the right and down.  This illustrates the auto-force effect (\ref{auto_forca}) between the charges, as discussed in section 2. From the point of view of traditional classical mechanics, it is as if the electromagnetic force between the charges behaves like an external force, leading to a non-conservation for the total mechanical momentum of the two-particle system.

\begin{figure}
[!htb]
\begin{center}
\hspace*{-10mm}\includegraphics[width= 17.0cm]{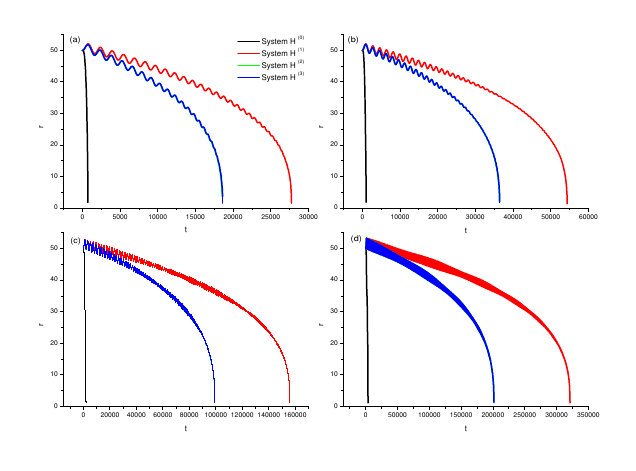}
\end{center}
\vspace*{-15mm}
\caption{\footnotesize The panels in figure show the time evolution of the relative distance between the particles for the trajectories obtained from the systems ${\bf H}^{(n)}$ ($n=0,1,2,3)$. Each panel corresponds for the following value of $\eta$: $\eta=1$ - panel (a), 
$\eta=2$ - panel (b), $\eta=5$ - panel (c) and $\eta=10$ - panel (d).}
\label{pai5}
\end{figure}

The main characteristic to observe the trajectories convergence is related with the singularity time. This can be seen from the evolution for the particles relative distance  shown in figure \ref{pai5}. Note that the relative distance evolution for the systems ${\bf H}^{(2)}$ (blue line) and ${\bf H}^{(3)}$ (cyan line) are practically coincident.
The singularity times, approximately estimated by  $t^{(n)}$, are shown in table 1 for different values of $\eta$.
Note that the times $t^{(2)}$ and $t^{(3)}$, respectively in the last two columns of the table, are very close to each other.
\begin{table}
[!htb]
\caption{The table shows the total integration times $t^{(n)}$ for systems defined by ${\bf H}^{(n)}$ with $n=1,2,3,4$ and $\eta= 1,2,5,10$. The total integration time is the time for the lightest charge to reach a speed of $0.8$, that is, $80\%$ of the speed of light. Note how the times $t^{(2)}$ (third column) and $t^{(3)}$ (fourth column) are virtually identical.} 
\begin{center}
\footnotesize
\begin{tabular}{llllllllllllll}
\br
 &&&  $t^{(0)}$ &&&  $t^{(1)}$ &&&   $t^{(2)}$ &&& $t^{(3)}$  \\
\mr
$\eta=1$ &&& $720.9$ &&&  $27740$ &&&   $18570$ &&& $18610$ \\
$\eta=2$ &&& $1003$ &&&  $54340$ &&&   $36380$ &&& $36430$ \\
$\eta=5$ &&& $1911$ &&&  $155900$ &&&   $99230$ &&& $99270$ \\
$\eta=10$ &&& $3496$ &&&  $321000$ &&&   $201400$ &&& $201400$\\
\br
\end{tabular}
\end{center}
\label{tabela1}
\end{table}

\begin{figure}
[!htb]
\begin{center}
\hspace*{-0mm}\includegraphics[width= 17.0cm]{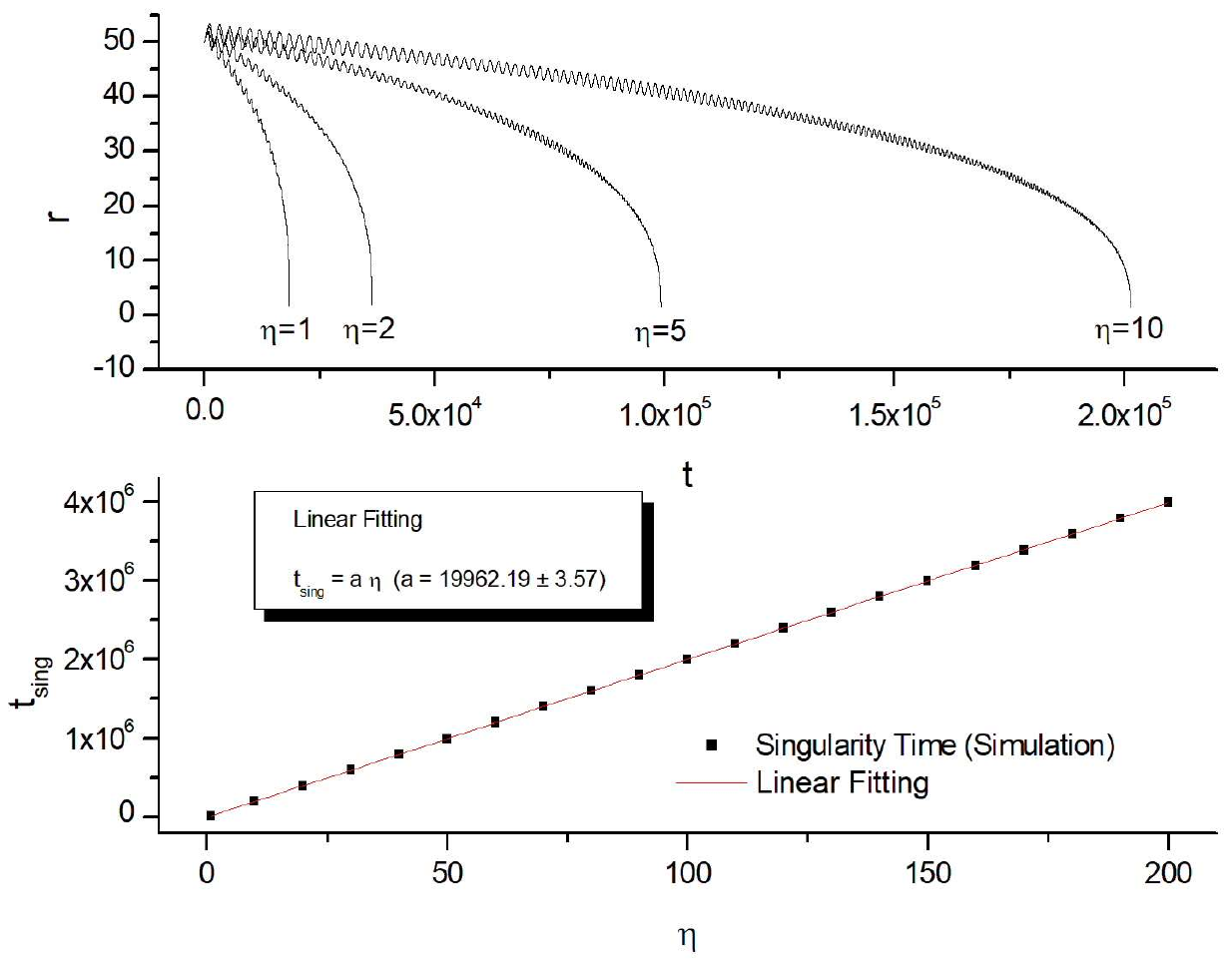}
\end{center}
\vspace*{-15mm}
\caption{\footnotesize The upper panel shows the time evolution of the relative distance $r=||{\bf r}_2-{\bf r}_1||$ for $\eta=1$, $\eta=2$, $\eta=5$ and $\eta=10$ using the trajectories obtained from the system ${\bf H}^{(3)}$. The bottom panel shows the time of singularity
$t_{sing}$ as a function of $\eta$. The singularity time was estimate as being approximately given by $t^{(3)}$, that is, the time spent by the fastest particle to become higher than $0.8$}  
\label{pai6}
\end{figure}

Figure \ref{pai6} shows the singularity time $t_{sing}$, estimated from the value obtained for $t^{(3)}$, as a function of $\eta$. We see that this relation can be very well adjusted by a linear relation, which is in accordance with the theoretical dependence obtained by Synge \cite{synge1940}, although in this work only an approximation for such  relation is calculated under the condition that $ \eta>>1$.

\subsection{Symmetric Electromagnetic Fields - $\alpha=0$}

For symmetric attractive fields, the trajectories are obtained through numerical integration of systems ${\bf H}^{(n)}$ ($n=0,1,2,3)$ considering $\alpha=0$. Let us remember that the initial conditions are the same used for $\alpha=1/2$.

\begin{figure}
[!htb]
\begin{center}
\hspace*{-10mm}\includegraphics[width= 17.0cm]{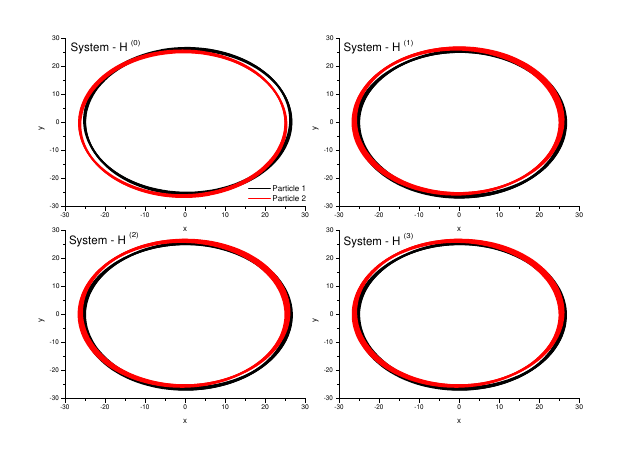}
\end{center}
\vspace*{-15mm}
\caption{\footnotesize Planar trajectories in Cartesian coordinates $(x,y)$ for the vectors ${\bf r}_1$, ${\bf r}_2$, where we have integrate numerically (for the same initial condition) the first four ODEs systems of the sequence (\ref{sequenciaHn}) considering  $\alpha=0$ and 
$\eta=1$. We can observe how the trajectories of the ${\bf H}^{(2)}$ system are practically identical to that of the ${\bf H}^{(3)}$ system}
\label{pai7}
\end{figure}

\begin{figure}
[!htb]
\begin{center}
\hspace*{-10mm}\includegraphics[width= 17.0cm]{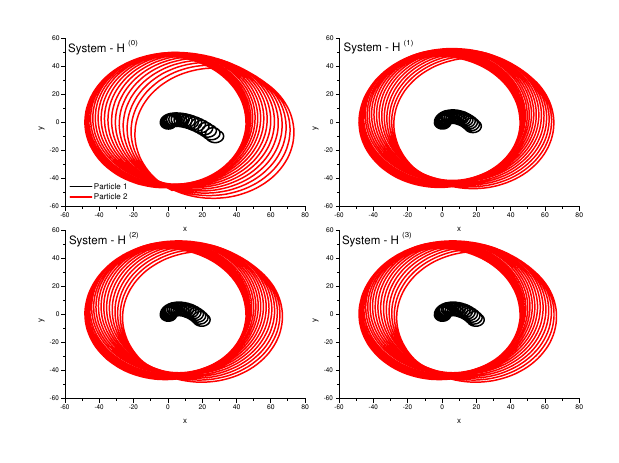}
\end{center}
\vspace*{-15mm}
\caption{\footnotesize Planar trajectories in Cartesian coordinates $(x,y)$ for the vectors ${\bf r}_1$, ${\bf r}_2$, where we have integrate numerically (for the same initial condition) the first four ODEs systems of the sequence (\ref{sequenciaHn}) considering $\alpha=0$ and 
$\eta=10$. We can observe how the trajectories of the ${\bf H}^{(2)}$ system are practically identical to that of the ${\bf H}^{(3)}$ system}
\label{pai8}
\end{figure}

Figures \ref{pai7} and \ref{pai8} show the trajectories of particles $1$ and $2$, respectively for  $\eta=1$ and $\eta=10$. We can see, as in the case of retarded fields, that the trajectories for the systems ${\bf H}^{(2)}$ and ${\bf H}^{(3)}$ are practically the same, evidencing the method convergence.
However, there is not  a singularity time  in which the particles collide. Moreover, the trajectories for the systems ${\bf H}^{(0)}$ and ${\bf H}^{(1 )}$ are closer to the trajectory of the system ${\bf H}^{(3)}$. We can also observe in figure \ref{pai8} the global displacement of the two charges trajectories  to the right, showing the effect of non-conservation of the total mechanical moment due to the self-force described in Eq. (\ref{auto_forca}).

For a given vector ${\bf f}$,  we denote ${\bf f}^{(n)}(t)$ as the curve of this vector  obtained by integrating the system ${\bf H}^{(n)}$.
Then, we define
\begin{eqnarray}
\label{distancia2}
D_{\bf f}[n,n+1]&=&\frac{1}{t_{max}}\int_0^{t_{max}}{\left|\left|{\bf f}^{(n+1)}(t)-{\bf f}^{(n)}(t)\right|\right|}dt. 
\end{eqnarray}
Let us  remember that  each  system ${\bf H}^{(n)}$ is integrated with the same initial condition and up to the same arbitrary maximum time $t_{max}$.
The numerical estimates of (\ref{distancia2}) for the particle position vectors are summarized in Table 2 below. In this table we can see evidence for the convergence of the trajectories according to the Cauchy criterion, that is, the distance between the trajectories obtained from the systems defined by the vector fields ${\bf H}^{(n+1)}$ and ${\bf H}^{(n)}$ seems to converge to zero as $n$ increases. We see that the distances $D_{{\bf r}_i}[2,3]$ are approximately three or four orders of magnitude smaller than the distances $D_{{\bf r}_i}[0,1]$ for $i=1,2$.

\begin{table}
[!htb]
\caption{The table shows the distances $D_{{\bf r}_i}[n,n+1]$ ($i=1,2$), as defined in Eq. (\ref{distancia2}), for $n=0,1,2,3$ and $\eta=1,2,5,10$.
Note, in the second and third columns, how these distances decrease as the value of $n$ increases.}
\begin{center}
\footnotesize
\begin{tabular}{lllll}
\mr
 &&  $D_{{\bf r}_1}[0,1]=1.379\times 10^1$ &&  $D_{{\bf r}_2}[0,1]=1.379\times 10^1$ \\
$\eta=1$ &&  $D_{{\bf r}_1}[1,2]=2.108\times 10^0$  && $D_{{\bf r}_2}[1,2]=2.108\times 10^0$ \\
 &&  $D_{{\bf r}_1}[2,3]=3.580\times 10^{-2}$ && $D_{{\bf r}_2}[2,3]=3.580\times 10^{-2}$  \\
\mr
&&  $D_{{\bf r}_1}[0,1]=7.802\times 10^0$  && $D_{{\bf r}_2}[0,1]=1.340\times 10^1$\\
$\eta=2$ &&  $D_{{\bf r}_1}[1,2]=1.050\times 10^0$ && $D_{{\bf r}_2}[1,2]=1.993\times 10^{0}$ \\
 &&  $D_{{\bf r}_1}[2,3]=1.982\times 10^{-2}$ && $D_{{\bf r}_2}[2,3]=3.761\times 10^{-2}$ \\
\mr
&&  $D_{{\bf r}_1}[0,1]=6.200\times 10^0$  && $D_{{\bf r}_2}[0,1]=9.516\times 10^0$ \\
$\eta=5$ &&  $D_{{\bf r}_1}[1,2]=6.340\times 10^{-1}$ && $D_{{\bf r}_2}[1,2]=1.124\times 10^{0}$\\
 &&  $D_{{\bf r}_1}[2,3]=5.276\times 10^{-3}$  && $D_{{\bf r}_2}[2,3]=1.184\times 10^{-2}$ \\
\mr
&&  $D_{{\bf r}_1}[0,1]=4.472\times 10^0$ && $D_{{\bf r}_2}[0,1]=5.667\times 10^0$ \\
$\eta=10$ &&  $D_{{\bf r}_1}[1,2]=7.232\times 10^{-1}$ && $D_{{\bf r}_2}[1,2]=8.057\times 10^{-1}$ \\
 &&  $D_{{\bf r}_1}[2,3]=1.607\times 10^{-3}$  && $D_{{\bf r}_2}[2,3]=1.131\times 10^{-2}$ \\
\mr
\end{tabular}
\end{center}  
\label{tabela2}
\end{table}

\begin{figure}
[!htb]
\begin{center}
\hspace*{-10mm}\includegraphics[width= 17.0cm]{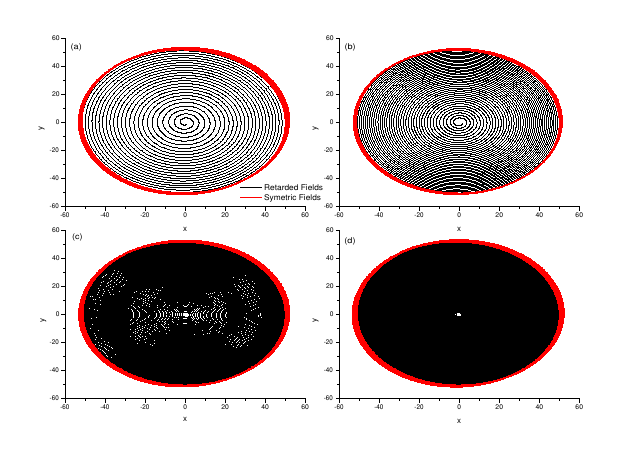}
\end{center}
\vspace*{-15mm}
\caption{\footnotesize Planar trajectories in Cartesian coordinates $(x,y)$ for the relative vector ${\bf r}={\bf r}_2-{\bf r}_1$, where we have integrate numerically (for the same initial condition) the ODEs systems defined by ${\bf H}^{(3)}$ in  the sequence (\ref{sequenciaHn}).
The respective value of $\eta$ in the panels are: $\eta=1$ (panel a), $\eta=2$ (panel b), $\eta=5$ (panel c) and $\eta=10$ (panel d).}
\label{pai9}
\end{figure}

\begin{figure}
[!htb]
\begin{center}
\hspace*{-10mm}\includegraphics[width= 17.0cm]{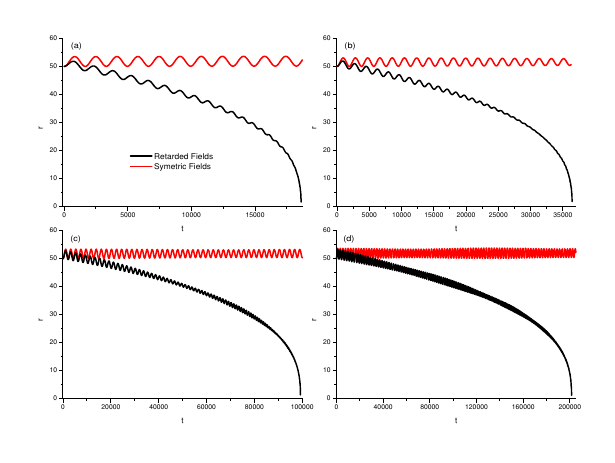}
\end{center}
\vspace*{-15mm}
\caption{\footnotesize Times evolution for the relative distance $r=||{\bf r}||=||{\bf r}_2-{\bf r}_1||$ corresponding to the respective trajectories showed in figure \ref{pai9}.
The respective value of $\eta$ in the panels are: $\eta=1$ (panel a), $\eta=2$ (panel b), $\eta=5$ (panel c) and $\eta=10$ (panel d).}
\label{pai10}
\end{figure}

We close this section by comparing the trajectories of  relative coordinates  obtained for systems with retarded fields ($\alpha=1/2$) with those obtained for systems with
symmetric fields ($\alpha=0$). We use the numerical integrations made with the ODEs systems defined by the vector field ${\bf H}^{(3)}$ for  $\eta=1$, $\eta=2 $, $\eta=5$ and $\eta=10$.
Figure \ref{pai9} shows the time evolution for the relative vector ${\bf r}={\bf r}_2-{\bf r}_1$ and figure \ref{pai10} for the respective module $r= ||{\bf r}||$, which stands for the relative distance between the particles.

We  see from figures \ref{pai9} and \ref{pai10} that the relative trajectories are quasi-circular:  in the case of retarded fields the quasi-circles spiral until the particles collide; in the case of symmetrical fields the quasi-circles (non-elliptical orbits) have radii that oscillate between a minimum and maximum value.

\section{Concluding Remarks}

The main result of this work was to develop a method to build a sequence of systems of ODEs, defined by vector fields ${\bf H}^{(n)}$
$(n=0,1,\ldots,\infty)$ according to Eq. (\ref{sistemaHn}), such that for a given initial condition on the positions and velocities of the charges, if the solution of the $n$th system of ODEs exists in a time interval $[t_0,t_1]$, with the sequence of these solutions converging to a limit function when $n\rightarrow\infty$, then this limit function will be part of a global solution of Synge's FDEs system defined in the interval $[t_0,t_1]$.

The main point of the method is to use the solution of the ODEs system defined by ${\bf H}^{(n)}$ to approximate the retarded and advanced times in order to  obtain the respective variables calculated at these times. We then use the approximations in Synge's FDEs system to define the next vector field ${\bf H}^{(n+1)}$ of the sequence. This sequence is initiated by the ODEs system defined by ${\bf H}^{(0)}$, which is obtained considering the instantaneous approximation for Synge's FDEs system.

This method allows the definition of a numerical integration algorithm for each  system of ODEs defined by the sequence of vector fields ${\bf H}^{(n)}$, and we can use any method for the numerical integration of systems of ODEs. In this work, we use the traditional fourth and fifth order Runge-Kutta methods. The drawbacks, in terms of computational efficiency, is that to compute the vector field ${\bf H}^{(n)}({\bf X})$ for a state vector ${\bf X} $, we have to recursively compute the previous vector fields ${\bf H}^{(i)}({\bf X})$ for $i=0,1,\ldots,n-1$.
In this algorithm, we update the trajectories of the charges integrating the same system of ODEs and, unlike Synge's method, we do not fix the trajectory of one of the charges to obtain the system of ODEs for the other charge. Indeed, Synge's algorithm does not concomitantly update the trajectory of the two charges and their convergence (if it exists) strictly depend on the fact that the masses of both charges can not be equal, which is not required by our method.

Another major problem in Synge's algorithm is having to completely reconstruct the trajectory of one of the charges at each step, once the trajectory of the other is fixed, making it difficult to control a finite time interval to coordinate the construction of each trajectory. In other words, to avoid convergence problems we should fix each trajectory in its maximal interval, however we do not know this maximal interval before calculating the trajectory.
The method developed in this work avoids this problem and can be considered as a kind of algorithm that allows the construction of a global solution from local solutions, that is, from a certain initial condition we can extend the solutions for the systems of ODEs defined by ${\bf H}^{(n)}$ up to their respective maximum intervals. In fact, as can be seen through the integration of systems with retarded fields ($\alpha=1/2$), the maximum interval (in the future) is defined by the singularity time $t_{sing}$ and we can go on extending our solution until this time. This can be seen in table \ref{tabela1}, where each vector field ${\bf H}^{(n)}$ has a different singularity time $t^{(n)}$, but these times converge to the time $t_{sing}$ as $n$ increases.
The extension of the solution to past times can be done without problem, since for each system of ODEs  we can extend the solutions to their maximal intervals, both in the future and in the past direction.

Originally, Synge's problem concerned only to the presence of retarded fields $(\alpha=1/2)$, but extending our algorithm to any combination of retarded and advanced fields does not imply any additional difficulty. Indeed, we extend the definition of Synge systems to encompass all possible combinations of retarded and advanced fields, since the distinctive factor of Synge's problems  is the issue of obtaining global solutions that are compatible with electromagnetic fields generated only by the point charges considered in the problem. In other words,  Synge's problem is the question about the existence of trajectories for two point charges isolated from the presence of external forces, regardless of whether they are produced by sources of electromagnetic or non-electromagnetic fields.

It is worth remembering that for Synge systems, the charge is a point without structure, that is, it does not take into account the self-force corrections, according to the Dirac-Lorentz renormalization process. The more general definition for these types of systems allows considering the Wheeler-Feymann equations as being a particular case of Synge systems. A very significant particular case in which the Dirac-Lorentz renormalization does not change the Synge system, since the self-force due to advanced fields cancel out those due to retarded fields.

If we consider general Dirac-Lorentz systems with linear combinations of retarded and advanced fields, as we did for Synge systems, then the Wheeler-Feymann equations define the unique problem of two isolated point charges that is at the same time a Synge and a Dirac-Lorentz system. For any non-symmetric combination of fields ($\alpha\neq 0$), the Dirac-Lorentz equations are different from the Synge equations and the charge trajectories, for each of these cases, will be different. In this work, we illustrate trajectories only for systems with retarded or symmetric fields, as these are the cases of greatest interest for physics.

The development of a method for obtaining global solutions for Dirac-Lorentz systems can be done in a equivalent way to what was developed for Synge systems.
The difference is that the sequence of vector fields ${\bf H}^{(n)}$ will define systems of first order ODEs, obtained from systems of third order FDEs. Thus, Cauchy's problem of finding a global solution will imply to know the initial values for the charge accelerations, which can be fixed independently of their positions and velocities. The algorithm development for Dirac-Lorentz systems and the comparison of  their global solutions with those obtained for Synge systems will be the subject of a future work.

\section{Acknowledgements}
This work was partially supported by CNPQ and CAPES (Federal Brazilian agencies).

\clearpage

\appendix

\section{Retarded Fields}

In this appendix we present the definitions of vectors and matrices, calculated at the respective retarded times, which appear in equation (\ref{a1-delay}). These vectors and matrices can be obtained directly from equation (\ref{eqG1}).

\noindent $\bullet$  More convenient definition of  vectors that point the direction of  fields:

\begin{equation} 
\hspace{-25mm}{\mathbf e}^{-}_{1}=  \mathbf{n}^-_{12} =\frac{\mathbf{r}_2(t)-\mathbf{r}_1(t^1_r)}{|\mathbf{r}_2(t)-\mathbf{r}_1(t^1_r)|}=
\left(\begin{array}{c}
e^-_{1x}\\e^-_{1y}\\e^-_{1z}
\end{array}\right),\;\;
{\mathbf e}^{-}_{2}=\mathbf{n}^-_{21}=\frac{\mathbf{r}_1(t)-\mathbf{r}_2(t^2_r)}{|\mathbf{r}_1(t)-\mathbf{r}_2(t^2_r)|}= 
\left(\begin{array}{c}
e^-_{2x}\\e^-_{2y}\\e^-_{2z}
\end{array}\right).
\end{equation}

\noindent $\bullet$  Definition of forces that are independent of accelerations:

\[
\hspace{-25mm}{\mathbf F}^{-}_1\hspace*{-1.5mm}=\hspace*{-1.5mm} 
\frac{S\left(1-{\mathbf v}^-_2\cdot{\mathbf v}^-_2\right)}{\left(1-{\mathbf v}^{-}_2\cdot{\mathbf e}^{-}_2\right)^3|{\mathbf r}_1-{\mathbf r}^-_2|^2}
\hspace*{-2mm}\left(\hspace*{-2mm}\begin {array}{c} { v_{1y}} 
\left({ v^-_{2x}}{ e^-_{2y} -{ v^-_{2y}}{ e^-_{2x}}}\right) 
+{ v_{1z}} 
\left({{ v^-_{2x}}{ e^-_{2z}- v^-_{2z}}{ e^-_{2x}}}\right) 
-{ v^-_{2x}}+{ e^-_{2x}}\\ \noalign{\medskip}{ v_{1x}} \left( { v^-_{2y}}{ e^-_{2x}-{ v^-_{2x}}{ e^-_{2y}}} \right) +{ v_{1z}} 
\left({ v^-_{2y}}{ e^-_{2z}-{ v^-_{2z}}{ e^-_{2y}}}\right) 
-{ v^-_{2y}}+{ e^-_{2y}}\\ \noalign{\medskip}{ v_{1x}} \left({ v^-_{2z}}{ e^-_{2x}-{ v^-_{2x}}{ e^-_{2z}}} \right) +{ v_{1y}} \left({ v^-_{2z}}{e^-_{2y}-{ v^-_{2y}}{ e^-_{2z}}} \right)-{ v^-_{2z}}+{e^-_{2z}}\end{array}\hspace*{-2mm} \right),
\]

\[
\hspace{-25mm}{\mathbf F}^{-}_2\hspace{-1.5mm}=\hspace{-1.5mm} 
\frac{S\left(1-{\mathbf v}^-_1\cdot{\mathbf v}^-_1\right)}{\left({\mathbf v}^{-}_{1}\cdot{\mathbf e}^{-}_{1}-1\right)^3|{\mathbf r}_2-{\mathbf r}^-_1|^2}
\hspace*{-2mm}\left(\hspace*{-2mm}\begin {array}{c} { v_{2y}} 
\left({ v^-_{1y}}{ e^-_{1x}} -{ v^-_{1x}}{ e^-_{1y}}\right) 
+{ v_{2z}}\left({ v^-_{1z}}{ e^-_{1x}} -{ v^-_{1x}}{ e^-_{1z}}\right)+{ v^-_{1x}}-{ e^-_{1x}}\\ \noalign{\medskip}{ v_{2x}} \left( { v^-_{1x}}{ e^-_{1y}}-{ v^-_{1y}}{ e^-_{1x}} \right) +{ v_{2z}} 
\left({ v^-_{1z}}{ e^-_{1y}} -{ v^-_{1y}}{ e^-_{1z}}\right) 
+{ v^-_{1y}}-{ e^-_{1y}}\\ \noalign{\medskip}{ v_{2x}} \left( { v^-_{1x}}{ e^-_{1z}}-{ v^-_{1z}}{ e^-_{1x}} \right) +{ v_{2y}} \left( { v^-_{1y}}{ e^-_{1z}}-{ v^-_{1z}}{ e^-_{1y}} \right) +{ v^-_{1z}}-{ e^-_{1z}}\end{array} \hspace*{-2mm}\right).
\]

\noindent $\bullet$ Definition of matrices that appear in forces that are dependent of accelerations:

\[
\hspace{-15mm}{\bf M}^-_{12}={\displaystyle \frac{S}{\left(1-{\mathbf v}^{-}_2\cdot{\mathbf e}^{-}_2\right)^3|{\mathbf r}^-_2-{\mathbf r}_1|}}{\bf L}^{\hspace*{-1mm}-}_{12},\,\,
{\bf M}^-_{21}=\frac{S}{\left({\mathbf v}^{-}_{1}\cdot{\mathbf e}^{-}_{1}-1\right)^3|{\mathbf r}_2-{\mathbf r}^-_1|}{\bf L}^{\hspace*{-1mm}-}_{21},
\]
with the elements of the matrix ${\bf L}^{\hspace*{-1mm}-}_{12}$ given by:
\begin{eqnarray}
\hspace*{0mm}{\left({\bf L}^{\hspace*{-1mm}-}_{12}\right)_{11}}=(-{ v^-_{2z}}{ e^-_{2y}}+{ v^-_{2y}}{ e^-_{2z}})(-{ v_{1y}}{ e^-_{2z}}+{ v_{1z}}{ e^-_{2y}}) \nonumber\\
\hspace*{43mm} + ({ v^-_{2y}}-{ e^-_{2y}})  ({ v_{1y}}-{ e^-_{2y}}) + ( { v^-_{2z}}-{ e^-_{2z}})  ( { v_{1z}}-{ e^-_{2z}})\nonumber
\end{eqnarray}
\[
\displaystyle {\left({\bf L}^{\hspace*{-1mm}-}_{12}\right)_{12}}= ( -{ v^-_{2x}}{ e^-_{2z}}+{ v^-_{2z}}{ e^-_{2x}})( -{ v_{1y}}{ e^-_{2z}}+{ v_{1z}}{ e^-_{2y}}) \\
\mbox{}- ( { v^-_{2x}}-{ e^-_{2x}})( { v_{1y}}-{ e^-_{2y}}) 
\]
\[
\displaystyle { \left({\bf L}^{\hspace*{-1mm}-}_{12}\right)_{13}} = ( -{ v^-_{2y}}{ e^-_{2x}}+{ v^-_{2x}}{ e^-_{2y}})(-{ v_{1y}}{ e^-_{2z}}+{ v_{1z}}{ e^-_{2y}}) \\
\mbox{}-( { v^-_{2x}}-{ e^-_{2x}})( { v_{1z}}-{ e^-_{2z}}) 
\]
\[
\displaystyle { \left({\bf L}^{\hspace*{-1mm}-}_{12}\right)_{21}}= (-{ v^-_{2y}}{ e^-_{2z}}+{ v^-_{2z}}{ e^-_{2y}})( -{ v_{1x}}{ e^-_{2z}}+{ v_{1z}}{ e^-_{2x}}) \\
\mbox{}-( { v^-_{2y}}-{ e^-_{2y}})({ v_{1x}}-{ e^-_{2x}}) 
\]
\begin{eqnarray}
\hspace*{0mm} {({\bf L}^{\hspace*{-1mm}-}_{12})_{22}}=(-{ v^-_{2z}}{ e^-_{2x}} +{ v^-_{2x}}{ e^-_{2z}})( -{ v_{1x}}{ e^-_{2z}}+{ v_{1z}}{ e^-_{2x}}) \nonumber\\
\hspace*{43mm} + ( { v^-_{2x}}-{ e^-_{2x}}) ( { v_{1x}}-{ e^-_{2x}}) + ( { v^-_{2z}}-{ e^-_{2z}}) ( { v_{1z}}-{ e^-_{2z}})\nonumber
\end{eqnarray}
\[
{ \left({\bf L}^{\hspace*{-1mm}-}_{12}\right)_{23}}= ( -{ v^-_{2x}}{ e^-_{2y}}+{ v^-_{2y}}{ e^-_{2x}})( -{ v_{1x}}{ e^-_{2z}}+{ v_{1z}}{ e^-_{2x}}) \\
\mbox{}-( { v^-_{2y}}-{ e^-_{2y}})( { v_{1z}}-{ e^-_{2z}}) 
\]
\[
{ \left({\bf L}^{\hspace*{-1mm}-}_{12}\right)_{31}}=
( { -v^-_{2z}}{ e^-_{2y}} +{ v^-_{2y}}{ e^-_{2z}}) (-{ v_{1x}}{ e^-_{2y}}+{ v_{1y}}{ e^-_{2x}}) \\
\mbox{}-({ v^-_{2z}}-{ e^-_{2z}})({ v_{1x}}-{ e^-_{2x}}) 
\]
\[
{ \left({\bf L}^{\hspace*{-1mm}-}_{12}\right)_{32}}= (-{ v^-_{2x}}{ e^-_{2z}}+{ v^-_{2z}}{ e^-_{2x}})(-{ v_{1x}}{ e^-_{2y}}+{ v_{1y}}{ e^-_{2x}}) \\
\mbox{}-({ v^-_{2z}}-{ e^-_{2z}})({ v_{1y}}-{ e^-_{2y}})
\]
\begin{eqnarray}
\hspace*{0mm}{({\bf L}^{\hspace*{-1mm}-}_{12})_{33}}=(-{ v^-_{2y}}{ e^-_{2x}} +{ v^-_{2x}}{ e^-_{2y}})( -{ v_{1x}}{ e^-_{2y}}+{ v_{1y}}{ e^-_{2x}}) \nonumber\\
\hspace*{38mm}+( { v^-_{2y}}-{ e^-_{2y}})  ( { v_{1y}}-{ e^-_{2y}}) +( { v^-_{2x}}-{ e^-_{2x}})  ({ v_{1x}}-{ e^-_{2x}});\nonumber
\end{eqnarray}
and the elements of the matrix ${\bf L}^{\hspace*{-1mm}-}_{21}$ given by
\begin{eqnarray}
\hspace*{0mm}\left({\bf L}^{\hspace*{-1mm}-}_{21}\right)_{11}= ( {v_{2y}}{e^-_{1z}}-{v_{2z}}{e^-_{1y}})( {v^-_{1y}}{e^-_{1z}}-{v^-_{1z}}{e^-_{1y}}) \nonumber \\
\hspace*{37mm}-( {v_{2y}}-{e^-_{1y}})( {v^-_{1y}}-{e^-_{1y}}) - ( {v_{2z}}-{e^-_{1z}})( {v^-_{1z}}-{e^-_{1z}})
\nonumber
\end{eqnarray}
\[\displaystyle \left({\bf L}^{\hspace*{-1mm}-}_{21}\right)_{12}= ( {v_{2z}}{e^-_{1y}}-{v_{2y}}{e^-_{1z}})( {v^-_{1x}}{e^-_{1z}}-{v^-_{1z}}{e^-_{1x}}) \\
\mbox{}+( {v_{2y}}-{e^-_{1y}})( {v^-_{1x}}-{e^-_{1x}})\]
\[\displaystyle \left({\bf L}^{\hspace*{-1mm}-}_{21}\right)_{13}=({v_{2y}}{e^-_{1z}}-{v_{2z}}{e^-_{1y}})( {v^-_{1x}}{e^-_{1y}}-{v^-_{1y}}{e^-_{1x}}) \\
\mbox{}+( {v_{2z}}-{e^-_{1z}})( {v^-_{1x}}-{e^-_{1x}})\]
\[\displaystyle \left({\bf L}^{\hspace*{-1mm}-}_{21}\right)_{21}=({v_{2z}}{e^-_{1x}}-{v_{2x}}{e^-_{1z}})( {v^-_{1y}}{e^-_{1z}}-{v^-_{1z}}{e^-_{1y}}) \\
\mbox{}+({v_{2x}}-{e^-_{1x}})( {v^-_{1y}}-{e^-_{1y}})\]
\begin{eqnarray}
\hspace*{0mm}\left({\bf L}^{\hspace*{-1mm}-}_{21}\right)_{22}=({v_{2x}}{e^-_{1z}}-{v_{2z}}{e^-_{1x}})({v^-_{1x}}{e^-_{1z}}-{v^-_{1z}}{e^-_{1x}}) \nonumber \\
\hspace*{36mm}-({v_{2x}}-{e^-_{1x}})( {v^-_{1x}}-{e^-_{1x}}) - ({v_{2z}}-{e^-_{1z}})({v^-_{1z}}-{e^-_{1z}})
\nonumber
\end{eqnarray}
\[\displaystyle \left({\bf L}^{\hspace*{-1mm}-}_{21}\right)_{23}= ({v_{2z}}{e^-_{1x}}-{v_{2x}}{e^-_{1z}})( {v^-_{1x}}{e^-_{1y}}-{v^-_{1y}}{e^-_{1x}}) \\
\mbox{}+( {v_{2z}}-{e^-_{1z}})({v^-_{1y}}-{e^-_{1y}})\]
\[\displaystyle \left({\bf L}^{\hspace*{-1mm}-}_{21}\right)_{31}=({v_{2x}}{e^-_{1y}}-{v_{2y}}{e^-_{1x}})({v^-_{1y}}{e^-_{1z}}-{v^-_{1z}}{e^-_{1y}}) \\ \mbox{}+({v_{2x}}-{e^-_{1x}})( {v^-_{1z}}-{e^-_{1z}})\]
\[\displaystyle \left({\bf L}^{\hspace*{-1mm}-}_{21}\right)_{32}=({v_{2y}}{e^-_{1x}}-{v_{2x}}{e^-_{1y}})( {v^-_{1x}}{e^-_{1z}}-{v^-_{1z}}{e^-_{1x}}) \\
\mbox{}+( {v_{2y}}-{e^-_{1y}})( {v^-_{1z}}-{e^-_{1z}})\]
\begin{eqnarray}
\hspace*{0mm}\left({\bf L}^{\hspace*{-1mm}-}_{21}\right)_{33}= ({v_{2x}}{e^-_{1y}}-{v_{2y}}{e^-_{1x}})( {v^-_{1x}}{e^-_{1y}}-{v^-_{1y}}{e^-_{1x}}) \nonumber \\
\hspace*{36mm}-( {v_{2x}}-{e^-_{1x}})( {v^-_{1x}}-{e^-_{1x}}) -( {v_{2y}}-{e^-_{1y}})( {v^-_{1y}}-{e^-_{1y}}).
\nonumber
\end{eqnarray}

\section{Advanced Fields}

In this appendix we present the definition of  vectors and matrices, calculated at the respective advanced times, which appear in equation (\ref{a1-delay}). These vectors and matrices can be obtained directly from equation (\ref{eqG1}).

\noindent $\bullet$ More convenient definition of  vectors that point the direction of  fields:

\begin{equation} 
\hspace{-25mm}{\mathbf e}^{+}_{1}=  \mathbf{n}^+_{12} =\frac{\mathbf{r}_2(t)-\mathbf{r}_1(t^a_2)}{|\mathbf{a}_2(t)-\mathbf{r}_1(t^a_1)|}=
\left(\begin{array}{c}
e^+_{1x}\\e^+_{1y}\\e^+_{1z}
\end{array}\right),\;\;
{\mathbf e}^{+}_{2}=\mathbf{n}^+_{21}=\frac{\mathbf{r}_1(t)-\mathbf{r}_2(t^a_2)}{|\mathbf{r}_1(t)-\mathbf{r}_2(t^a_2)|}= 
\left(\begin{array}{c}
e^+_{2x}\\e^+_{2y}\\e^+_{2z}
\end{array}\right).
\end{equation}

\noindent $\bullet$ Definition of forces that are independent of accelerations:

\[
\hspace{-25mm}{\mathbf F}^{+}_1\hspace*{-1.5mm}=\hspace*{-1.5mm} 
\frac{S\left(1-{\mathbf v}^+_2\cdot{\mathbf v}^+_2\right)}{\left(1+{\mathbf v}^{+}_2\cdot{\mathbf e}^{+}_2\right)^3|{\mathbf r}_1-{\mathbf r}^+_2|^2}
\hspace*{-2mm}\left(\hspace*{-2mm}\begin {array}{c} -{ v_{1y}} 
\left({ v^+_{2x}}{ e^+_{2y} -{ v^+_{2y}}{ e^+_{2x}}}\right) 
-{ v_{1z}} 
\left({{ v^+_{2x}}{ e^+_{2z}- v^+_{2z}}{ e^+_{2x}}}\right) 
+{ v^+_{2x}}+{ e^+_{2x}}\\ \noalign{\medskip}-{ v_{1x}} \left( { v^+_{2y}}{ e^+_{2x}-{ v^+_{2x}}{ e^+_{2y}}} \right) -{ v_{1z}} 
\left({ v^+_{2y}}{ e^+_{2z}-{ v^+_{2z}}{ e^+_{2y}}}\right) 
+{ v^+_{2y}}+{ e^+_{2y}}\\ \noalign{\medskip}-{ v_{1x}} \left({ v^+_{2z}}{ e^+_{2x}-{ v^+_{2x}}{ e^+_{2z}}} \right) -{ v_{1y}} \left({ v^+_{2z}}{e^-_{2y}-{ v^+_{2y}}{ e^+_{2z}}} \right)+{ v^+_{2z}}+{e^-_{2z}}\end{array}\hspace*{-2mm} \right),
\]

\[
\hspace{-25mm}{\mathbf F}^{+}_2\hspace{-1.5mm}=\hspace{-1.5mm} 
\frac{-S\left(1-{\mathbf v}^+_1\cdot{\mathbf v}^+_1\right)}{\left({\mathbf v}^{+}_{1}\cdot{\mathbf e}^{+}_{1}+1\right)^3|{\mathbf r}_2-{\mathbf r}^+_1|^2}
\hspace*{-2mm}\left(\hspace*{-2mm}\begin {array}{c} -{ v_{2y}} 
\left({ v^+_{1y}}{ e^+_{1x}} -{ v^+_{1x}}{ e^+_{1y}}\right) 
-{ v_{2z}}\left({ v^+_{1z}}{ e^+_{1x}} -{ v^+_{1x}}{ e^+_{1z}}\right)-{ v^+_{1x}}-{ e^+_{1x}}\\ \noalign{\medskip}-{ v_{2x}} \left( { v^+_{1x}}{ e^+_{1y}}-{ v^+_{1y}}{ e^+_{1x}} \right) -{ v_{2z}} 
\left({ v^+_{1z}}{ e^+_{1y}} -{ v^+_{1y}}{ e^+_{1z}}\right) 
-{ v^+_{1y}}-{ e^+_{1y}}\\ \noalign{\medskip}{ -v_{2x}} \left( { v^+_{1x}}{ e^+_{1z}}-{ v^+_{1z}}{ e^+_{1x}} \right) -{ v_{2y}} \left( { v^+_{1y}}{ e^+_{1z}}-{ v^+_{1z}}{ e^+_{1y}} \right) -{ v^+_{1z}}-{ e^+_{1z}}\end{array} \hspace*{-2mm}\right).
\]

\noindent $\bullet$  Definition of  matrices that appear in forces that are dependent of accelerations:

\[
\hspace{-15mm}{\bf M}^+_{12}={\displaystyle \frac{S}{\left(1+{\mathbf v}^{-}_2\cdot{\mathbf e}^{+}_2\right)^3|{\mathbf r}^+_2-{\mathbf r}_1|}}{\bf L}^{\hspace*{-1mm}+}_{12},\,\,
{\bf M}^+_{21}=\frac{-S}{\left({\mathbf v}^{+}_{1}\cdot{\mathbf e}^{+}_{1}+1\right)^3|{\mathbf r}_2-{\mathbf r}^+_1|}{\bf L}^{\hspace*{-1mm}+}_{21},
\]
with the elements of the matrix ${\bf L}^{\hspace*{-1mm}+}_{12}$ given by:
\begin{eqnarray}
\hspace*{0mm}{\left({\bf L}^{\hspace*{-1mm}+}_{12}\right)_{11}}=({ v^+_{2z}}{ e^+_{2y}}-{ v^+_{2y}}{ e^+_{2z}})(-{ v_{1y}}{ e^+_{2z}}+{ v_{1z}}{ e^+_{2y}}) \nonumber\\
\hspace*{39mm}  -({ v^+_{2y}}+{ e^+_{2y}})  ({ v_{1y}}-{ e^+_{2y}}) - ( { v^+_{2z}}+{ e^+_{2z}})  ( { v_{1z}}-{ e^+_{2z}})\nonumber
\end{eqnarray}
\[
\displaystyle {\left({\bf L}^{\hspace*{-1mm}+}_{12}\right)_{12}}= ({ v^+_{2x}}{ e^+_{2z}}-{ v^+_{2z}}{ e^+_{2x}})( -{ v_{1y}}{ e^+_{2z}}+{ v_{1z}}{ e^+_{2y}}) \\
\mbox{}+( { v^+_{2x}}+{ e^+_{2x}})( { v_{1y}}-{ e^+_{2y}}) 
\]
\[
\displaystyle { \left({\bf L}^{\hspace*{-1mm}+}_{12}\right)_{13}} = ( { v^+_{2y}}{ e^+_{2x}}-{ v^+_{2x}}{ e^+_{2y}})(-{ v_{1y}}{ e^+_{2z}}+{ v_{1z}}{ e^+_{2y}}) \\
\mbox{}+( { v^+_{2x}}+{ e^+_{2x}})( { v_{1z}}-{ e^+_{2z}}) 
\]
\[
\displaystyle { \left({\bf L}^{\hspace*{-1mm}+}_{12}\right)_{21}}= ({ v^+_{2y}}{ e^+_{2z}}-{ v^+_{2z}}{ e^+_{2y}})( -{ v_{1x}}{ e^+_{2z}}+{ v_{1z}}{ e^+_{2x}}) \\
\mbox{}+( { v^+_{2y}}+{ e^+_{2y}})({ v_{1x}}-{ e^+_{2x}}) 
\]
\begin{eqnarray}
\hspace*{0mm} { ({\bf L}^{\hspace*{-1mm}+}_{12})_{22}}=({ v^+_{2z}}{ e^+_{2x}} -{ v^+_{2x}}{ e^+_{2z}}) ( -{ v_{1x}}{ e^+_{2z}}+{ v_{1z}}{ e^+_{2x}} ) \nonumber\\
\hspace*{39mm} - ( { v^+_{2x}}+{ e^+_{2x}}) ( { v_{1x}}-{ e^+_{2x}} ) - ( { v^+_{2z}}+{ e^+_{2z}} )  ( { v_{1z}}-{ e^+_{2z}} )\nonumber
\end{eqnarray}
\[
{ \left({\bf L}^{\hspace*{-1mm}+}_{12}\right)_{23}}= ({ v^+_{2x}}{ e^+_{2y}}-{ v^+_{2y}}{ e^+_{2x}})( -{ v_{1x}}{ e^+_{2z}}+{ v_{1z}}{ e^+_{2x}}) \\
\mbox{}+( { v^+_{2y}}+{ e^+_{2y}})( { v_{1z}}-{ e^+_{2z}}) 
\]
\[
{ \left({\bf L}^{\hspace*{-1mm}+}_{12}\right)_{31}}=
( { v^+_{2z}}{ e^+_{2y}} -{ v^+_{2y}}{ e^+_{2z}}) (-{ v_{1x}}{ e^+_{2y}}+{ v_{1y}}{ e^+_{2x}}) \\
\mbox{}+({ v^+_{2z}}-{ e^+_{2z}})({ v_{1x}}-{ e^+_{2x}}) 
\]
\[
{ \left({\bf L}^{\hspace*{-1mm}+}_{12}\right)_{32}}= ({ v^+_{2x}}{ e^+_{2z}}-{ v^+_{2z}}{ e^+_{2x}})(-{ v_{1x}}{ e^+_{2y}}+{ v_{1y}}{ e^+_{2x}}) \\
\mbox{}+({ v^+_{2z}}+{ e^+_{2z}})({ v_{1y}}-{ e^+_{2y}})
\]
\begin{eqnarray}
\hspace*{0mm}{ \left({\bf L}^{\hspace*{-1mm}+}_{12}\right)_{33}}=({ v^+_{2y}}{ e^+_{2x}} -{ v^+_{2x}}{ e^+_{2y}})( -{ v_{1x}}{ e^+_{2y}}+{ v_{1y}}{ e^+_{2x}}) \nonumber\\
\hspace*{39mm}-( { v^+_{2y}}+{ e^+_{2y}})  ( { v_{1y}}-{ e^+_{2y}}) -( { v^+_{2x}}+{ e^+_{2x}})  ({ v_{1x}}-{ e^+_{2x}});\nonumber
\end{eqnarray}
and the elements of the matrix ${\bf L}^{\hspace*{-1mm}+}_{21}$ given by
\begin{eqnarray}
\hspace*{0mm}\left({\bf L}^{\hspace*{-1mm}+}_{21}\right)_{11}= ( {v_{2y}}{e^+_{1z}}-{v_{2z}}{e^+_{1y}})( -{v^+_{1y}}{e^+_{1z}}+{v^+_{1z}}{e^+_{1y}}) \nonumber \\
\hspace*{40mm}+( {v_{2y}}-{e^+_{1y}})( {v^+_{1y}}+{e^+_{1y}}) + ( {v_{2z}}-{e^+_{1z}})( {v^+_{1z}}+{e^+_{1z}})
\nonumber
\end{eqnarray}
\[\displaystyle \left({\bf L}^{\hspace*{-1mm}+}_{21}\right)_{12}= ( {v_{2z}}{e^+_{1y}}-{v_{2y}}{e^+_{1z}})( -{v^+_{1x}}{e^+_{1z}}+{v^+_{1z}}{e^+_{1x}}) \\
\mbox{}-( {v_{2y}}+{e^+_{1y}})( {v^+_{1x}}+{e^+_{1x}})
\]
\[\displaystyle \left({\bf L}^{\hspace*{-1mm}+}_{21}\right)_{13}=({v_{2y}}{e^+_{1z}}-{v_{2z}}{e^+_{1y}})( -{v^+_{1x}}{e^+_{1y}}+{v^+_{1y}}{e^+_{1x}}) \\
\mbox{}-( {v_{2z}}-{e^+_{1z}})( {v^+_{1x}}+{e^+_{1x}})
\]
\[
\displaystyle \left({\bf L}^{\hspace*{-1mm}+}_{21}\right)_{21}=({v_{2z}}{e^+_{1x}}-{v_{2x}}{e^+_{1z}})( -{v^+_{1y}}{e^+_{1z}}+{v^+_{1z}}{e^+_{1y}}) \\
\mbox{}-({v_{2x}}-{e^+_{1x}})( {v^+_{1y}}+{e^+_{1y}})
\]
\begin{eqnarray}
\hspace*{0mm}\left({\bf L}^{\hspace*{-1mm}+}_{22}\right)_{22}=({v_{2x}}{e^+_{1z}}-{v_{2z}}{e^+_{1x}})(-{v^+_{1x}}{e^+_{1z}}+{v^+_{1z}}{e^+_{1x}}) \nonumber \\
\hspace*{40mm}+({v_{2x}}-{e^+_{1x}})( {v^+_{1x}}+{e^+_{1x}}) +({v_{2z}}-{e^+_{1z}})({v^+_{1z}}+{e^+_{1z}})
\nonumber
\end{eqnarray}
\[
\displaystyle \left({\bf L}^{\hspace*{-1mm}+}_{21}\right)_{23}= ({v_{2z}}{e^+_{1x}}-{v_{2x}}{e^+_{1z}})( -{v^+_{1x}}{e^+_{1y}}+{v^+_{1y}}{e^+_{1x}}) \\
\mbox{}-( {v_{2z}}-{e^+_{1z}})({v^+_{1y}}+{e^+_{1y}})
\]
\[
\displaystyle \left({\bf L}^{\hspace*{-1mm}+}_{21}\right)_{31}=({v_{2x}}{e^+_{1y}}-{v_{2y}}{e^+_{1x}})(-{v^+_{1y}}{e^+_{1z}}+{v^+_{1z}}{e^+_{1y}}) \\ \mbox{}-({v_{2x}}-{e^+_{1x}})( {v^+_{1z}}+{e^+_{1z}})
\]
\[
\displaystyle \left({\bf L}^{\hspace*{-1mm}+}_{21}\right)_{32}=({v_{2y}}{e^+_{1x}}-{v_{2x}}{e^+_{1y}})( -{v^+_{1x}}{e^+_{1z}}+{v^+_{1z}}{e^+_{1x}}) \\
\mbox{}-( {v_{2y}}-{e^+_{1y}})( {v^+_{1z}}+{e^+_{1z}})
\]
\begin{eqnarray}
\hspace*{0mm}\left({\bf L}^{\hspace*{-1mm}+}_{21}\right)_{33}= ({v_{2x}}{e^+_{1y}}-{v_{2y}}{e^+_{1x}})( -{v^+_{1x}}{e^+_{1y}}+{v^+_{1y}}{e^+_{1x}}) \nonumber \\
\hspace*{40mm}+( {v_{2x}}-{e^+_{1x}})( {v^+_{1x}}+{e^+_{1x}}) +( {v_{2y}}-{e^+_{1y}})( {v^+_{1y}}+{e^+_{1y}}).
\nonumber
\end{eqnarray}

\section{Formulas for bidimensional (planar) system}

 In order to obtain the planar system from Eq. (\ref{a1-delay}), let us consider that the vectors in equation (\ref{vetores}) are represented in a cartesian plane $(x,y)$. Then, if we define
\begin{eqnarray} 
{\bf e}^-_{1}(t_1^r) =\frac{{\bf r}_2-{\bf r}_1(t_1^r)}{|{\bf r}_2-{\bf r}_1(t_1^r)|}=
\left(\begin{array}{c}
 e^-_{1x}\\ e^-_{1y}
\end{array}\right),\;\;
{\bf e}^-_2(t_2^r)
=\frac{{\bf r}_1-{\bf r}_2(t_2^r)}{|{\bf r}_1-{\bf r}_2(t_2^r)|}= 
\left(\begin{array}{c}
 e^-_{2x}\\ e^-_{2y}
\end{array}\right),\nonumber
\end{eqnarray}
\begin{eqnarray} 
{\bf e}^+_{1}(t_1^a) =\frac{{\bf r}_2-{\bf r}_1(t_1^a)}{|{\bf r}_2-{\bf r}_1(t_1^a)|}=
\left(\begin{array}{c}
 e^+_{1x}\\ e^+_{1y}
\end{array}\right),\;\;
{\bf e}^+_2(t_2^a)
=\frac{{\bf r}_1-{\bf r}_2(t_2^a)}{|{\bf r}_1-{\bf r}_2(t_2^a)|}= 
\left(\begin{array}{c}
 e^+_{2x}\\ e^+_{2y}
\end{array}\right),\nonumber
\end{eqnarray}
we can obtain the following expressions for the vectors and matrices (given in Appendices A and B) that appear in  equation (\ref{a1-delay-antigo}):
$$
{\bf F}^{-}_1(t_2^r)= 
\frac{S\left(1-{\mathbf v}_2(t^r_2)\cdot{\mathbf v}_2(t^r_2)\right)}{\left(1-{\mathbf v}_2(t^r_2)\cdot{\mathbf e}^-_2(t^r_2)\right)^3\left|{\mathbf r}_1-{\mathbf r}_2(t^r_2)\right|^2}
\hspace*{-2mm}\left(\hspace*{-2mm}
\begin {array}{c} v_{1y}\hspace{-1mm} 
\left[v_{2x}(t_2^r){ e^-_{2y}- v_{2y}(t_2^r){ e^-_{2x}}}\right]  
-v_{2x}(t_2^r)+{ e^-_{2x}}\\ 
\noalign{\medskip}
v_{1x}\hspace*{-1mm} \left[v_{2y}(t_2^r){ e^-_{2x}- v_{2x}(t_2^r){ e^-_{2y}}} \right] 
-v_{2y}(t_2^r)+{ e^-_{2y}}
\end{array}\hspace*{-2mm}\right),
$$
$$
{\bf F}^{+}_1(t_2^a)= 
\frac{S\left(1-{\mathbf v}_2(t^a_2)\cdot{\mathbf v}_2(t^a_2)\right)}{\left(1+{\mathbf v}_2(t^a_2)\cdot{\mathbf e}^+_2(t^a_2)\right)^3\left|{\mathbf r}_1-{\mathbf r}_2(t^a_2)\right|^2}
\hspace*{-2mm}\left(\hspace*{-2mm}
\begin {array}{c} v_{1y}\hspace{-1mm} 
\left[v_{2y}(t_2^a){e^+_{2x}-v_{2x}(t_2^a){e^+_{2y}}}\right]  
+v_{2x}(t_2^a)+{  e^+_{2x}}\\ 
\noalign{\medskip}
v_{1x}\hspace*{-1mm} \left[ v_{2x}(t_2^a){  e^+_{2y}-v_{2y}(t_2^a){  e^+_{2x}}} \right] 
+v_{2y}(t_2^a)+{  e^+_{2y}}
\end{array}\hspace*{-2mm}\right),
$$
$$
{\bf F}^{-}_2(t_1^r)= 
\frac{S\left(1-{\mathbf v}_1(t^r_1)\cdot{\mathbf v}_1(t^r_1)\right)}{\left({\mathbf v}_{1}(t^r_1)\cdot{\mathbf e}^-_{1}(t^r_1)-1\right)^3\left|{\mathbf r}_2-{\mathbf r}_1(t^r_1)\right|^2}
\hspace*{-2mm}\left(\hspace*{-2mm}
\begin {array}{c} { v_{2y}}\hspace*{-1mm} 
\left[{ v_{1y}}(t_1^r){ e^-_{1x}} -{ v_{1x}}(t_1^r){ e^-_{1y}}\right]
+{ v_{1x}}(t_1^r)-{ e^-_{1x}}\\ 
\noalign{\medskip}
{ v_{2x}}\hspace*{-1mm} \left[ { v_{1x}}(t_1^r){ e^-_{1y}}-{ v_{1y}}(t_1^r){ e^-_{1x}} \right] 
+{ v_{1y}}(t_1^r)-{ e^-_{1y}}
\end{array} \hspace*{-2mm}\right),
$$
$$
{\bf F}^{+}_2(t_1^a)= 
\frac{-S\left(1-{\mathbf v}_1(t^a_1)\cdot{\mathbf v}_1(t^a_1)\right)}{\left({\mathbf v}_{1}(t^a_1)\cdot{\mathbf e}^+_{1}(t^a_1)+1\right)^3\left|{\mathbf r}_2-{\mathbf r}_1(t^a_1)\right|^2}
\hspace*{-2mm}\left(\hspace*{-2mm}
\begin {array}{c} { v_{2y}}\hspace*{-1mm} 
\left[{ v_{1x}}(t_1^a){  e^+_{1y}-{ v_{1y}}(t_1^a){  e^+_{1x}}}\right]
-{ v_{1x}}(t_1^a)-{  e^+_{1x}}\\ 
\noalign{\medskip}
{ v_{2x}}\hspace*{-1mm} \left[{ v_{1y}}(t_1^a){  e^+_{1x}- { v_{1x}}(t_1^a){  e^+_{1y}}} \right] 
-{ v_{1y}}(t_1^a)-{  e^+_{1y}}
\end{array} \hspace*{-2mm}\right),
$$
\[\hspace*{-25mm}
{\bf M}^-_{12}(t^r_2)={\displaystyle \frac{S\hspace*{1mm}{\bf L}^{\hspace*{-1mm}-}_{12}(t^r_2)}{\left(1-{\mathbf v}_2(t^r_2)\cdot{{\mathbf e}}^{-}_2(t^r_2)\right)^3
\left|{\mathbf r}_1-{\mathbf r}_2(t^r_2)\right|}},
{\bf M}^-_{21}(t^r_1)=\frac{S\hspace*{1mm}{\bf L}^{\hspace*{-1mm}-}_{21}(t^r_2)}{\left({\mathbf v}_{1}(t^r_1)\cdot{\mathbf e}^{-}_{1}(t^r_1)-1\right)^3
\left|{\mathbf r}_2-{\mathbf r}_1(t^r_1)\right|},
\]
\[\hspace*{-25mm}
{\bf M}^+_{12}(t^r_2)={\displaystyle \frac{S\hspace*{1mm}{\bf L}^{\hspace*{-1mm}+}_{12}(t^a_2)}{\left(1+{\mathbf v}_2(t^a_2)\cdot{{\mathbf e}}^{+}_2(t^a_2)\right)^3
\left|{\mathbf r}_1-{\mathbf r}_2(t^a_2)\right|}},
{\bf M}^+_{21}(t^a_1)=\frac{-S\hspace*{1mm}{\bf L}^{\hspace*{-1mm}+}_{21}(t^a_2)}{\left({\mathbf v}_{1}(t^a_1)\cdot{\mathbf e}^{+}_{1}(t^a_1)+1\right)^3
\left|{\mathbf r}_2-{\mathbf r}_1(t^a_1)\right|}.
\]
The matrices  ${\bf L}^{\hspace*{-1mm}-}_{12}(t^r_2)$,  ${\bf L}^{\hspace*{-1mm}+}_{12}(t^a_2)$, ${\bf L}^{\hspace*{-1mm}-}_{21}(t^r_1)$ and ${\bf L}^{\hspace*{-1mm}+}_{21}(t^a_1)$ are defined as
$$
{\bf L}^{\hspace*{-1mm}-}_{12}(t_2^r)=
\left(
\begin{array}{cc}
\left[{ v_{2y}}(t_2^r)-{ e^-_{2y}}\right] \left[{ v_{1y}}-{ e^-_{2y}}\right] & - \left[{ v_{2x}}(t_2^r)-{ e^-_{2x}}\right]\left[{ v_{1y}}-{ e^-_{2y}}\right] \\
-\left[{ v_{2y}}(t_2^r)-{ e^-_{2y}}\right]\left[{ v_{1x}}-{ e^-_{2x}}\right] &  \left[{ v_{2x}}(t_2^r)-{ e^-_{2x}}\right]\left[{ v_{1x}}-{ e^-_{2x}}\right]
\end{array}\right),
$$
$$
{\bf L}^{\hspace*{-1mm}+}_{12}(t_2^a)=
\left(
\begin{array}{cc}
-\left[{ v_{2y}}(t_2^a)+{ e^+_{2y}}\right] \left[{ v_{1y}}-{ e^+_{2y}}\right] & \left[{ v_{2x}}(t_2^a)+{ e^+_{2x}}\right]\left[{ v_{1y}}-{ e^+_{2y}}\right] \\
\left[{ v_{2y}}(t_2^a)+{ e^+_{2y}}\right]\left[{ v_{1x}}-{ e^+_{2x}}\right] &  -\left[{ v_{2x}}(t_2^a)+{ e^+_{2x}}\right]\left[{ v_{1x}}-{ e^+_{2x}}\right]
\end{array}\right),
$$
$$
{\bf L}^{\hspace*{-1mm}-}_{21}(t_1^r)=
\left(\begin{array}{cc}
-\left[{v_{2y}}-{ e^-_{1y}}\right]\left[{v_{1y}}(t_1^r)-{ e^-_{1y}}\right] & \left[{v_{2y}}-{ e^-_{1y}}\right]\left[{v_{1x}}(t_1^r)-{ e^-_{1x}}\right] \\ 
\left[{v_{2x}}-{ e^-_{1x}}\right]\left[{v_{1y}}(t_1^r)-{ e^-_{1y}}\right] & -\left[{v_{2x}}-{ e^-_{1x}}\right]\left[ {v_{1x}}(t_1^r)-{ e^-_{1x}}\right]
\end{array}\right),
$$
$$
{\bf L}^{\hspace*{-1mm}+}_{21}(t_1^a)=
\left(\begin{array}{cc}
\left[{v_{2y}}-{ e^+_{1y}}\right]\left[{v_{1y}}(t_1^a)+{ e^+_{1y}}\right] & -\left[{v_{2y}}-{ e^+_{1y}}\right]\left[{v_{1x}}(t_1^a)+{ e^+_{1x}}\right] \\ 
-\left[{v_{2x}}-{ e^+_{1x}}\right]\left[{v_{1y}}(t_1^a)+{ e^+_{1y}}\right] & \left[{v_{2x}}-{ e^+_{1x}}\right]\left[ {v_{1x}}(t_1^a)+{ e^+_{1x}}\right]
\end{array}\right).
$$
The matrices ${\cal M}^{-1}_{11}$ and ${\cal M}^{-1}_{22}$ in Eq. ({\ref{matrizesMii}), evaluated at the present time $t$,  depends on the projector operators defined as follows:
\begin{eqnarray}
&{\cal Q}_{\hat{\bf v}_1}=
\left(\begin{array}{cc}
{\hat v}_{1x}^2 & {\hat v}_{1x}{\hat v}_{1y} \\
{\hat v}_{1y}{\hat v}_{1x}(t) & {\hat v}_{1y}^2 
\end{array}\right),\;\;
{\cal Q}_{\hat{\bf v}_2}=
\left(\begin{array}{cc}
{\hat v}_{2x}^2 & {\hat v}_{2x}{\hat v}_{2y} \\
{\hat v}_{2y}{\hat v}_{2x} & {\hat v}_{2y}^2 
\end{array}\right),& \nonumber\\
&\displaystyle \hat{\bf v}_1=\frac{{\bf v}_1}{\left|{\bf v}_1\right|}=
\left(
\begin{array}{c}
{\hat v}_{1x}\\{\hat v}_{1y}
\end{array}
\right),\,\,
\displaystyle \hat{\bf v}_2=\frac{{\bf v}_2}{\left|{\bf v}_2\right|}
=\left(\begin{array}{c}
{\hat v}_{2x}\\{\hat v}_{2y}
\end{array}\right).& \nonumber
\end{eqnarray}

\end{document}